\documentclass[aps, prb, reprint, amsmath, amssymb, floatfix]{revtex4-2}

\usepackage{graphicx}
\usepackage{dcolumn}
\usepackage{txfonts} 
\usepackage{bm}
\usepackage{xcolor}
\usepackage[colorlinks, citecolor=blue, linkcolor=blue]{hyperref}

\begin{document}

\title{Complex dynamics of nano-oscillators with dual vortex free layers mutually coupled via spin-torques}
\author{Loukas Kokkinos}
\email{loukas.kokkinos@universite-paris-saclay.fr}
\author{Joo-Von Kim}
\email{joo-von.kim@universite-paris-saclay.fr}
\affiliation{Centre de Nanosciences et de Nanotechnologies, CNRS, Université Paris-Saclay, 91120 Palaiseau, France
}

\date{\today}

\begin{abstract}
Spin-torque vortex oscillators (STVOs) have been shown to exhibit rich and complex dynamical regimes, which are strongly dependent on the polarizer's configuration. Here, we give an overview of the dynamics in an STVO comprising two vortex free layers, where each layer serves as a dynamic polarizer for the other, increasing the number of degrees of freedom and therefore, the complexity of the system. The dynamics are studied through extensive micromagnetic simulations, performed using our own implementation of the coupled equations of motion, implemented in the open-source micromagnetics code \texttt{MuMax3}. We explore the roles of relative vortex configurations and layer asymmetry on the current-driven dynamics, and find several complex regimes, including self-modulated gyration, the emergence of C-state dynamics, as well as chaotic transitions between regular gyration and this C-state. 

\end{abstract}
\maketitle

\section{Introduction}
Spin-torque nano-oscillators (STNOs) comprise magnetoresistive multilayers in which spin-polarized currents induce self-sustained magnetization oscillations. These arise when the intrinsic dissipation mechanisms are overcome by the torques~\cite{slonczewski_current-driven_1996, berger_emission_1996} associated with the flow of spin angular momentum supplied by spin-polarized currents traversing the multilayer, resulting in steady-state magnetization precession that can span a wide range of frequencies in the microwave regime depending on the oscillation mode excited~\cite{kiselev_microwave_2003, rippard_direct_2004, pribiag_magnetic_2007, mistral_current_2008, garciasanchez_skyrmion_2016, wittrock_beyond_2021}. Because of their frequency tunability, STNOs have drawn significant interest for potential applications in telecommunications~\cite{villard_ghz_2010, litvinenko_analog_2021} and as microwave field sources for magnetic storage~\cite{zhu_microwave_2008, zhu_microwave_2010}. More recently, great strides have also been made in exploring the use of STNOs for unconventional computing, such as neuro-inspired paradigms~\cite{torrejon_neuromorphic_2017, romera_vowel_2018, riou_temporal_2019, imai_associative_2023}, stochastic computing~\cite{phan_unbiased_2024}, and chaos-based information processing~\cite{yoo_pattern_2020}, owing to the vast range of complex dynamical regimes they can access.

Oscillators based on vortex dynamics within magnetoresistive nanopillars~\cite{pribiag_magnetic_2007, khvalkovskiy_vortex_2009, dussaux_large_2010}, termed spin-torque vortex oscillators (STVOs), have been the subject of much focus due to their high quality factor and large signal output in magnetic tunnel junctions~\cite{dussaux_large_2010}. In these geometries, the magnetic free layer comprises a thin-film disk whose aspect ratio favors the magnetic vortex as an equilibrium ground state. In general, a minimum requirement for the polarizer layer of the spin valve or magnetic tunnel junction is to possess a component perpendicular to the film plane, which is often induced by an applied perpendicular magnetic field. Within reduced variable models like the Thiele equation~\cite{ivanov_excitation_2007, kim_spintorque_2012}, it can be shown that this perpendicular component results in an effective force that counteracts the viscous damping of the vortex motion, thereby allowing steady-state gyration of the vortex core about the disk center to be sustained. The constraint of a perpendicular polarizer can be relaxed if the magnetic configuration of the polarizer is nonuniform. For example, it has been shown theoretically and through numerical simulations that a polarizer in a vortex state can provide substantial torques to sustain vortex gyration in the free layer~\cite{khvalkovskiy_nonuniformity_2010}. Nonuniform in-plane polarizers can also trigger transitions between the usual gyration and self-sustained oscillations of the `C-state'~\cite{wittrock_beyond_2021}, which represents a curling magnetization configuration in the shape of the letter `C'. This state can be considered as an extension of the vortex state in which the core resides virtually outside of the disk.

Further complexity in the oscillator dynamics can arise if the usual assumption of a static polarizer configuration is relaxed. As Slonczewski discussed in his seminal paper~\cite{slonczewski_current-driven_1996}, back-scattered electrons from the free layer can also induce a spin-torque on the magnetization in the polarizer layer, thereby resulting in an additional coupling between the two layers through mutual spin-torques. This is usually ignored in real devices as the polarizer typically comprises a synthetic antiferromagnet that is exchange-biased by a metallic antiferromagnet. When this constraint is lifted, such as in ``dual free-layer'' systems, coupled dynamics can appear in which the rotation of the magnetization in one free layer will subsequently pull the other, resulting in a perpetual, staggered rotation of the magnetizations that resemble the motion of windmill blades. Dual free layers with perpendicular magnetic anisotropy exhibit complex switching dynamics~\cite{matsumoto_chaos_2019, farcis_spiking_2023}, which can be detected as spikes in the magnetoresistance, or more complex waveforms resulting from synchronisation and chaos~\cite{taniguchi_synchronized_2019}. Despite several studies on vertically-coupled vortices through dipole interactions~\cite{cherepov_core-core_2012,koop_2014_nonlin,bondarenko_2019_chaos}, only few have focused on vortices coupled by mutual spin torques~\cite{locatelli_dynamics_2011, sluka_quenched_2012,hamadeh_diverse_2025}.

In this article, we examine in detail the dynamics of dual free layer STVOs. Through extensive micromagnetic simulations, we discuss the role of the relative vortex configurations (chirality, polarity) on the current-driven dynamics. We also explore the role of layer asymmetry and find regimes in which chaotic transitions between gyration and C-state dynamics occur. We find that these transitions provide clear magnetoresistance signals and can be biased with perpendicular applied magnetic fields.

\section{Model \label{sec:model}}
\begin{figure}
\centering
	\includegraphics[width=8.5cm]{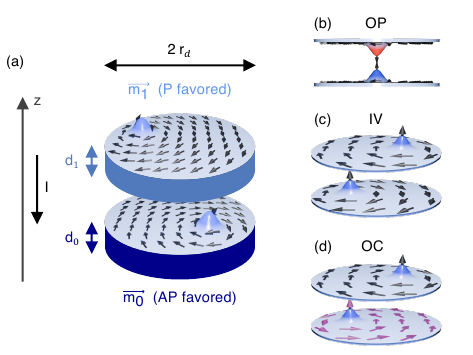}
	\caption{(a) Schematic of the two free layers for $p_{0}=p_{1}=$1 and $c_{0}=c_{1}=$-1. $d_0$, $d_1$ and $r_\mathrm{d}$ are the layers dimensions. $I$ is the current passing through the oscillator. Layer 0 (1) favors an antiparallel (parallel) configuration in respect to layer 1 (0) due to STT. (b), (c) and (d) show the three studied relative vortex configurations. (b) opposite polarities configuration with $p_{0}=1$, $p_{1}=-1$ and $c_{0}=c_{1}=$-1. (c) identical vortices configuration with $p_{0}=p_{1}=-1$ and $c_{0}=c_{1}=$-1. (d) opposite chiralities configuration with $p_{0}=p_{1}=-1$ and $c_{0}=1$, and $c_{1}=-1$.}  
	\label{fig:System}
\end{figure}

We consider the dual free-layer STVO system shown in Fig.~\ref{fig:System}(a).
It comprises two ferromagnetic, disks with identical radius $r_d$ and film thickness $d_i$, where $i=0,1$ denote the layer number. The conventional current is taken to flow in the $-z$ direction, perpendicular to the film plane, which corresponds to the flow of electrons in the $+z$ direction. The dynamics of the magnetization in each layer is governed by the Landau-Lifshitz equation with Gilbert damping and mutual spin-transfer torques,
\begin{subequations}
\begin{align}
    \frac{\partial \mathbf{m}_0}{\partial t} &= \mathbf{m}_0 \times \left( -|\gamma_0| \mathbf{H}_{\mathrm{eff},0} + \alpha \frac{\partial \mathbf{m}_0}{\partial t} -  \beta_0 \mathbf{m}_0 \times \mathbf{m}_1 \right), \\
    \frac{\partial \mathbf{m}_1}{\partial t} &= \mathbf{m}_1 \times \left( -|\gamma_0| \mathbf{H}_{\mathrm{eff},1} + \alpha \frac{\partial \mathbf{m}_1}{\partial t} + \beta_1 \mathbf{m}_1 \times \mathbf{m}_0 \right),
\end{align}
\label{eq:LLGS}
\end{subequations}
where $\gamma$ is the gyromagnetic ratio, $\mu_0$ the vacuum permeability, $\gamma_0 = \mu_0 \gamma$, and $\mathbf{m}(\mathbf{r},t)_i$ is a unit vector field ($\| \mathbf{m} \| = 1$) representing the magnetization configuration in layer $i$. The effective field,
\begin{equation}
	\mathbf{H}_{\mathrm{eff},i} = -\frac{1}{\mu_0 M_s} \frac{\delta U}{\delta \mathbf{m}_i},
\end{equation}
represents the variational derivative of the total magnetic energy, $U$, with respect to the magnetization. $U$ comprises the exchange, dipole-dipole, and Zeeman interactions. Besides static applied fields, the latter also includes contributions from the Oersted-Amp{\`e}re field related to the flow of the current $I$ through the nanopillar, which is given by
\begin{equation}
    \mathbf{H}_\mathrm{Oe}(x,y) = \frac{I}{2\pi r^{2}_{d}}  (y\mathbf{e}_{x} - x \mathbf{e}_{y}),
    \label{eq:Boe}  
\end{equation}
where $(x,y)$ denote the Cartesian coordinates within the film plane. The current density is assumed to be uniform across the disk. $\alpha$ represents the Gilbert damping constant and $\beta_i$ represents the strength of spin-torques on layer $i$,
\begin{equation} 
	\beta_i = \frac{g \mu_B}{2 M_s (\pi r_d^2 d)} \frac{P I}{|e|},
    \label{eq:betaSTT}
\end{equation}
where $g$ is the gyromagnetic ratio, $\mu_B$ is the Bohr magneton, $e$ is the electron charge, $P$ is the spin polarization, and $M_{s,i}$ is the layer-dependent saturation magnetization. We note that the magnetizations within layers $0$ and $1$ are coupled together through dipolar interactions (which are accounted for in $H_\mathrm{eff}$) and the mutual spin-torque term proportional to $\beta$.

We implemented the coupled equations of motion in Eq.~(\ref{eq:LLGS}) in the open-source micromagnetics code \texttt{MuMax3}~\cite{vansteenkiste_design_2014}. The code performs numerical time integration of the Landau-Lifshitz equation with Gilbert damping and Slonczewski spin torques using the finite difference method. Our implementation extends the standard version by allowing for a dynamical, non-uniform polarizer and mutual spin-transfer torques between the two free layers. We verified our code against results obtained using the standard implementation for Slonczewski spin-transfer torques for a single free layer with fixed polarizer and with literature results for macrospin windmill dynamics~\cite{matsumoto_chaos_2019}. We used micromagnetic parameters consistent with permalloy, with a base value of the saturation magnetization of $M_{s} = 800$~kA/m, an exchange constant of $A = 13$ pJ/m, and a Gilbert damping constant of $\alpha = 0.01$. For the geometrical parameters, we take the disk radius to be $r_d = 250$~nm and free layer thicknesses of $d_{0} = d_{1} \equiv d  = 7$~nm. The nanopillar is discretized using $256 \times 256 \times 2$ finite difference cells; the spacer layer is ignored. The current flow is along the $-z$ direction as shown in Fig.~\ref{fig:System}, and is considered uniform throughout the whole disk. Within this configuration, spin-torques acting on $\mathbf{m}_0$ favor an antiparallel (AP) alignment with respect to $\mathbf{m}_1$, while at the same time the torques acting on $\mathbf{m}_1$ will favor a parallel (P) alignment with $\mathbf{m}_0$. For simplicity and without loss of generality, we take the current polarization to be $P=1$.

Since the spin-transfer torques have no effect when the magnetization of the two layers is strictly collinear, i.e., no torques appear if $\mathbf{m}_0(\mathbf{r}) \times \mathbf{m}_1(\mathbf{r}) = \mathbf{0}$ everywhere, dynamics can only be initiated if the two vortex cores do not fully overlap.  Displacing the cores from the disk center and from one another can be achieved by applying an in-plane static field of a few mT, in addition to the STT, over a few nanoseconds. In what follows, we will discuss results from simulations in which the initial configuration comprises vortices in both layers in which both cores are displaced from the disk center, along the $x$ axis, by $-60$~nm and 40~nm respectively.

\section{Role of relative vortex configurations  \label{sec:relativeConfiguration}}
We first focus on the role of the relative configurations of the vortices in the two layers. The vortex state is defined by two parameters, the polarity $p$, which is defined by the sign of the perpendicular magnetization $m_z$ at the core center, and the chirality $c$, which defines the sense of circulation of the magnetic moments within the film plane, with $c=-1$ ($c=1$) corresponding to clockwise (counter-clockwise) circulation when viewed from $+z$. In damped oscillations of the vortex core, the polarity determines the sense of gyration, with $p=1$ leading to counterclockwise gyration of the core about the disk center when viewed from $+z$. The sign of the core polarity can be switched using a perpendicular external field, whereas the chirality can be controlled by the Oersted-Amp{\`e}re field. For spin valves with vortex free layers, it has been shown that the chirality and polarity within each layer can be selected if one free layer is thicker than the other~\cite{locatelli_dynamics_2011}.

In this section, we will consider three distinct vortex configurations for free layers with identical thickness. The first involves the case of opposite polarity, $p_0 p_1 =-1$, and identical chirality, $c_0 = c_1 =-1$ [Fig.~\ref{fig:System}(b)], denoted as OP. The sign of the chirality chosen corresponds to the case where the magnetic moments exhibit the same circulation as the current-induced Oersted-Amp{\`e}re field. It has been shown experimentally that this case represents a favorable configuration for generating measurable radiofrequency (RF) oscillations~\cite{locatelli_dynamics_2011}, with subsequent studies also focused on this system~\cite{hamadeh_core_2024,hamadeh_diverse_2025}.  The second case involves the identical vortex (IV) configurations, with $p_0 = p_1 = 1$ and $c_0 = c_1 = -1$, as illustrated in Fig.~\ref{fig:System}(c). The third case is the opposite chirality case (OC), in which the polarities are identical, $p_0 = p_1 = 1$, whilst $c_0 = -c_1 = 1$, as shown in Fig.~\ref{fig:System}(d).

\begin{figure}
	\centering\includegraphics[width=8.5cm]{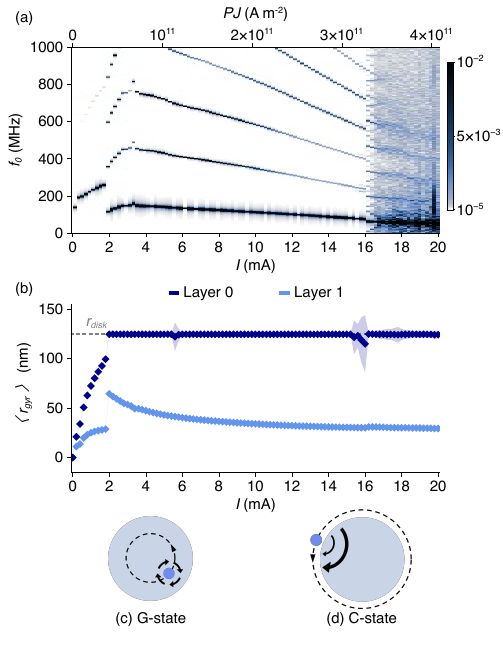}
	\caption{(a) Power spectral density map of the gyration frequency of the vortex in layer 0 as a function of the applied dc current $I$ in the case of opposite polarities. (b) Average gyration orbit radii as a function of the current applied to the system for layers 0 (dark) and 1 (light) are represented in solid markers. The standard deviation is also given in the form of a semi-transparent envelope for both layers. (c)G-state schematic, where the vortex core gyrates inside the disk. (d) C-state schematic, where the virtual vortex core gyrates outside the disk.}
	\label{fig:PSD_OP}
\end{figure}

The power spectrum of vortex dynamics in the OP case is presented in Fig.~\ref{fig:PSD_OP}(a), which shows how the power spectral density (PSD) of magnetization oscillations evolve as a function of applied current.
The PSD at each applied current is computed as follows. First, we initiate the system where the vortices in both layers are displaced from the center (Sec.~\ref{sec:model}) and let the system evolve over 500 ns, during which the average magnetization of both layers and the core positions are recorded. The PSD, $S(f)$, is then computed from the last $t_0 = $ 350 ns of the simulations from the spatially-averaged $m_x(t) = (1/V)\int m_x(\mathbf{r},t) \, dV$ component in layer 0, using the Hann windowing function $w(t)$ to minimize spectral leakage,
\begin{equation}
S(f) = \left| \int_{0}^{t_0} \, w(t) \, m_x(t) \, e^{-i 2 \pi f t} \, dt \right|^2,
\end{equation}
which is then represented as a color map in Fig.~\ref{fig:PSD_OP}(a).

We observe a strong response in the PSD as soon as a finite value of the current is applied. Due to their opposite polarities, the two cores initially undergo gyration in opposite directions (i.e., clockwise and counterclockwise). After a transient period, however, the motion becomes synchronised with the gyrotropic dynamics of layer 0 being dominant. Because of this synchronised dynamics, the PSD in Fig.~\ref{fig:PSD_OP}(a) is representative of the overall dynamics in both layers. In this coupled gyrotropic motion, we observe a phase difference $\delta \Phi = \pi$ between the two core positions, which is consistent with windmill dynamics. Within the low current regime of $I < 2$~mA, the gyration frequency increases as a function of current, which is typical of STVOs. A number of harmonics of the main gyration frequency can also be seen. At $I = 2$~mA, a sudden drop in frequency is observed, signalling a transition to a different dynamical mode. As the current is further increased, we observe a gradual, steady decrease in the gyration frequency. At large currents, $I > 16$~mA, the power spectrum exhibits a large level of athermal noise, suggesting a transition towards a chaotic state.

In order to better understand the different dynamical regimes at play, we examine how the time-averaged radius of gyration, $\langle r_{gyr}(t) \rangle$, evolves in both layers with the applied current, as shown in Fig.~\ref{fig:PSD_OP}(b). 

We compute its average over the same time interval as for the PSD map. Within the low current regime, $I < 2$~mA, the gyration radii in both layers increase with increasing current, with the largest changes taking place in layer $0$. Within this range, the vortex in each disk undergoes gyration about the disk center; we term this the G-state [Fig.~\ref{fig:PSD_OP}(c)]. At $I=2$~mA, the radius of gyration $r_\mathrm{gyr}$ in layer $0$ attains the value of the disk radius, $r_\mathrm{gyr} = r_\mathrm{d}$, indicating that the core is expelled from the disk. This corresponds to the transition toward the dynamical C-state~\cite{wittrock_beyond_2021},  in which the vortex core is expelled and gyrates virtually outside the disk, leading to the rotation of a C-shaped magnetization as illustrated in Fig.~\ref{fig:PSD_OP}(d). The C-state in layer 0 is observed for all currents $I>2$~mA, while the magnetization in layer 1 remains in the G-state. In this new \emph{windmill C-state regime}, we still have $\delta \Phi = \pi$ between the core in layer 1, and the virtual core in layer 0. In contrast to single free layer STVOs, this OP windmill system does not rely on strong perpendicular applied fields or high currents to enter the C-state. The current-dependence of the gyration frequency is also qualitatively different, as the frequency starts to decrease with current ($\partial f_0/\partial I < 0$) above $I=\textrm{3.4 mA}$.

\begin{figure}
\centering
	\includegraphics[width=8.5cm]{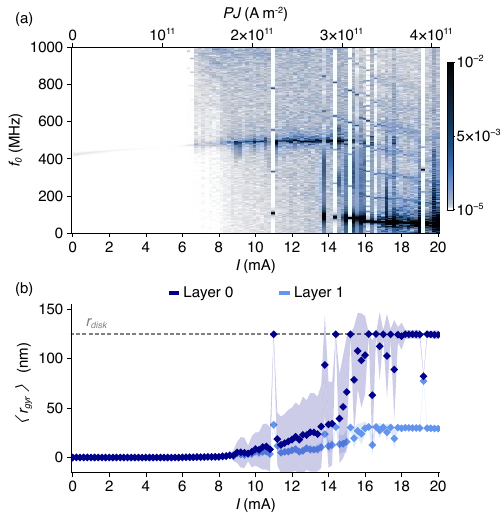}
	\caption{(a) Power spectral density map of the gyration frequency of the vortex in layer 0 as a function of the applied dc current $I$ in the case of identical vortices. (b) Average gyration orbit radii as a function of the current applied to the system for layers 0 (dark) and 1 (light) are represented in solid markers. The standard deviation is also given in the form of a semi-transparent envelope for both layers. } 
	\label{fig:PSD_IV}
\end{figure}

In Fig.~\ref{fig:PSD_IV}, we present the PSD map [Fig.~\ref{fig:PSD_IV}(a)] and the time-averaged radius of gyration  [Fig.~\ref{fig:PSD_IV}(b)] for identical vortices, computed in the same way presented for the OP case. Unlike the OP case, the IV configuration exhibits no dynamics for a wide range of currents.
Up to $I = 8$~mA, the two cores relax towards their equilibrium position at the center of the disk, $r_\mathrm{gyr,0} = r_\mathrm{gyr,1} = 0$. In this configuration, the dipolar coupling between the two cores will be attractive (as opposed to the repulsive one in the OP case), leading to their overlap and a subsequent reduction in the mutual spin torques. At high currents $I > 15$~mA, we observe the same dynamical behavior as in the OP case, where layer 0 enters the C-state. Between these two regions ($8 - 15$~mA), the gyration radii and frequencies are nonzero and show almost linear increase with current, with very strong fluctuations. These fluctuations suggest there exists some nonlinear mechanism that induces the gyration of the cores in this region, originating from layer 0, where the fluctuations are much stronger.

\begin{figure}
\centering
	\includegraphics[width=8.5cm]{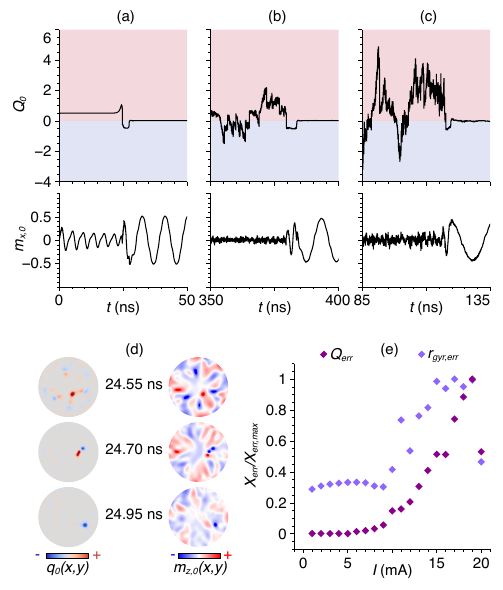}
	\caption{\label{fig:CGS transition} Topological charge, $Q_0$, and magnetization component along $x$ axis, $m_\mathrm{x,0}$, in layer 0 (bottom layer) as a function of time for fully identical vortices under applied currents of (a) $11$~mA, (b) $15$~mA, and (c) $19$~mA. (d) Profiles of the topological charge density $q_0 (x,y)$ (left) and $m_\mathrm{z,0}$ (right). At $t=24.55$~ns, vortex/antivortex pairs are generated in the disk, in addition to the original vortex core. At $t = 24.70$ ns the original core interacts with vortex antivortex pair. At 24.95 ns the original core and the antivortex annihilate, leaving a new core of opposite polarity. (e) Normalized standard deviations for $Q_0$ and $r_{gyr,0}$, computed over the first $125$~ns of the simulation, as a function of current. }  
\end{figure}

In order to understand this mechanism, we focus on the topological charge in layer 0,
\begin{equation}
  Q_{0} = \frac{1}{4\pi} \int \mathbf{m}_{0}(\mathbf{r}) \cdot \left (\frac{\partial \mathbf{m}_{0}}{\partial x} \times \frac{\partial \mathbf{m}_{0}}{\partial y} \right) dV,
\label{eq:topo_charge}
\end{equation}
which allows us to track the evolution of the topological magnetic textures inside this layer. In Figs.~\ref{fig:CGS transition}(a), \ref{fig:CGS transition}(b), and \ref{fig:CGS transition}(c), we present time traces of $Q_{0}$ for currents of 11, 15, and 19 mA, respectively. These traces also include the magnetic component along the $x$-axis, $m_{x,0}$. All simulations begin with $Q_{0}=0.5$, consistent with a vortex of positive core polarity.

For $I=11$~mA, $Q_{0}$ remains constant for the first 25~ns before abruptly dropping to $-0.5$. This transition signifies a reversal of the core, while the subsequent jump to $Q_{0}=0$ suggests the core is annihilated. Concurrently, $m_{x,0}$ transitions to oscillations of high amplitude and low frequency, characteristic of the C-state regime observed in the OP configuration. This behavior aligns with the core’s disappearance from the disk.

Similar core reversal and C-state transitions occur for $15$ and $19$~mA. These transitions are preceded by significant fluctuations in $Q_{0}$, indicating the emergence of topological structures, such as vortex/antivortex pairs. These pairs explain the transitions as the core interacts with one of them [Fig.~\ref{fig:CGS transition} (d)]. This interaction triggers a core reversal through a mechanism akin to that described in Ref.~\onlinecite{van_waeyenberge_magnetic_2006}. Consequently, the system recovers an OP configuration, and the core exits the disk as expected in this current range.

Figure~\ref{fig:CGS transition}(e) presents the behavior of the normalized standard deviations for topological charge, $Q_{err}$, and gyration radius, $r_{gyr,err}$, in layer 0, computed over the first 125~ns of simulations. Both sets of points exhibit a similar qualitative pattern: a nearly constant value until 9 mA, followed by an increase of similar slopes. Therefore, the number of emerging pairs increases with current, statistically decreasing the time for the core reversal to occur. These fluctuations causing the transition occur in layer 0, not layer 1, as a result of the mutual spin torques. While layer 1 favors a parallel configuration between the layers, which is satisfied as the cores begin to overlap, layer 0 pushes for antiparallel alignment. Consequently, above a certain current threshold, the magnetic moments in layer 0 become frustrated in this parallel configuration and attempt to anti-align, resulting in the observed fluctuations. The roles would be reversed if sign of the applied current is reversed.

\begin{figure}
\centering
	\includegraphics[width=8.5cm]{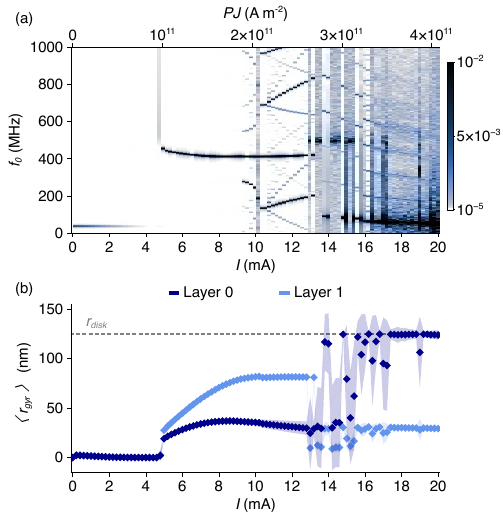}
	\caption{\label{fig4:PSD opposite chiralities} (a) Power spectral density map of the gyration frequency of the vortex in layer 0 as a function of the applied dc current $I$ in the case of opposite chiralities. (b) Average gyration orbit radii as a function of the current applied to the system for layers 0 (dark) and 1 (light) are represented in solid markers. The standard deviation is also given in the form of a semi-transparent envelope for both layers.} 
\end{figure}

Figure~\ref{fig4:PSD opposite chiralities} illustrates the dynamics for the opposite chirality configuration. Gyration of the cores begins at $I=5$ mA, and, as in the previous cases, a windmill C-state emerges around $13.6$~mA. However, the gyration regime where both layers are in the G-state exhibits significant differences. Unlike the OP and IV cases, $r_{gyr,1}>r_{gyr,0}$ between 5 and 12.8 mA, and the frequency remains nearly constant at around 410 MHz. At 13.6 mA, we observe $r_{gyr,0}>r_{gyr,1}$ again, but with a broader and noisier spectrum, similar to the IV case. This current represents a threshold above which the chirality in layer 0 switches to align with the Oersted field, transitioning the system into the IV configuration. The transition from the OC to the IV configuration also explains the shift to a C-state at higher currents. 
\begin{figure*}
\centering
	\includegraphics[width=17cm]{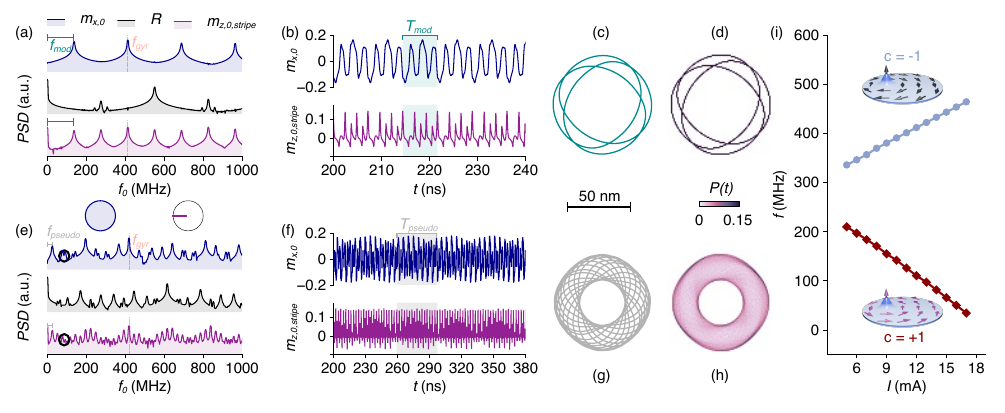}
	\caption{\label{fig:OC_mod} Commensurate (10.4 mA) and incommensurate (chaotic) (12.8 mA) regimes for OC modulated system. (a) and (e) show the frequency spectra of the layer 0 averaged $m_{x,0}$ (top), the magnetoresistance $R$ (middle) and of the $z$ component of the magnetization in a radial cut $m_{z,0,stripe}$ (bottom) for currents 10.4 and 12.8 mA respectively. In both plots is highlighted the gyration frequency of the cores around 410 MHz. In (a) is also highlighted the modulation frequency, $f_\mathrm{mod}$, and in (e) the pseudo-modulation frequency, $f_\mathrm{pseudo}$. (b) and (f) show time traces of $m_\mathrm{x,0}$ and $m_\mathrm{z,0,stripe}$ for currents 10.4 and 12.8 mA respectively. In (b) is highlighted the modulation period $T_\mathrm{mod}=1/f_\mathrm{mod}$. In (f) is highlighted the pseudo-modulation period $T_\mathrm{pseudo}=1/f_\mathrm{pseudo}$. (c) and (g) show the core trajectories in layer 0 over one (pseudo-)orbit at 10.4 and 12.8 mA respectively. (d) and (h) show density maps of the core position over 550 ns at 10.4 and 12.8 mA respectively (g) Relaxation eigen-frequencies of the vortex for positive (diamonds) and negative (cirlces) chirality in the presence of an Oersted field as function of the corresponding current.}
\end{figure*}

It is further observed that for currents between 9 and 14 mA, several narrow frequency bands exist for each current. In Figure~\ref{fig:OC_mod}(a), three frequency spectra are given, computed by Fourier transform over the last 350 ns of a 500 ns simulation at 10.4 mA. The top spectrum represents the Fourier transform of the disk-averaged $m_{x,0}$, which is used to create the map in Fig.~\ref{fig4:PSD opposite chiralities}(a). A frequency comb, with a peak separation of $2f_\mathrm{gyr}/3$, is formed around the gyration frequency, $f_\mathrm{gyr}$. This evidence suggests some self-modulation within the system. This modulation induces subharmonic dynamics, occurring at $f_\mathrm{gyr}/3$. As a pattern repeats every three periods of gyration in the $m_{x,0}$ time trace in Fig.~\ref{fig:OC_mod}(b) (top plot), the modulation translates into non-circular core trajectories. This is evident in Fig.~\ref{fig:OC_mod}(c) for layer 0, where the core orbit now resembles a hypotrochoid. The periodicity over three gyration periods means that the core returns to its initial position after three full turns ($2\pi$), explaining the peak at $f_\mathrm{gyr}/3$. The periodicity is further confirmed by the position density plot, $P(t)$, in Fig.~\ref{fig:OC_mod}(d), computed over 550 ns, where the trajectory is identical to the orbit over $T_\mathrm{mod}$. The middle spectrum in Fig.~\ref{fig:OC_mod}(a) is obtained by Fourier transform of the magnetoresistance, 
\begin{equation}
R = - \frac{1}{\pi r^{2}_{d}}\int  m_{0}(\mathbf{r}) \cdot m_{1}(\mathbf{r})\text{ } d\mathbf{r}^2,
\label{eq:MR}
\end{equation} 
over the the same time interval as the $m_{x,0}$ one. The resistance is a direct indicator of the relative position between the cores of the two layers, which explains the differences between the $m_{x,0}$ and $R$ spectra. One is the Fourier transform of one layer, while the other is the Fourier transform of the scalar product between magnetizations. At 10.4 mA, the main resistance frequency peak occurs at $f_\mathrm{res}=4f_\mathrm{gyr}/3$. This can be explained by the four-fold symmetry of the core orbits, as shown in Fig.~\ref{fig:OC_mod}(c). The two cores maintain a nearly constant phase difference and recover four times a position equivalent to their initial one over one full orbit. A modulation of $2f_\mathrm{gyr}/3$ can also be observed in the resistance spectrum, with sidebands being much weaker in amplitude in this case. The bottom spectrum in Fig.~\ref{fig:OC_mod}(a) is obtained by Fourier transform of $m_{z,0,stripe}$, the $z$ component of the magnetization along a radial stripe in layer 0. This cut is shown in the right schematic below the spectrum. It demonstrates that there is a local modulation of the magnetization with a frequency of $f_\mathrm{mod}=f_\mathrm{gyr}/3$. This local modulation is visible in the $m_{z,0,stripe}$ time trace in Fig.~\ref{fig:OC_mod}(b) and contributes to the larger-scale modulation of the core dynamics in the disks.

Figure~\ref{fig:OC_mod}(e) displays the $m_\mathrm{x,0}$ (top), $R$ (middle), and $m_\mathrm{z,0,stripe}$ (bottom) spectra computed using the same method for a current of $I=12.8$ mA. Similar to the case at 10.4 mA, local modulations produce a rich $m_{z,0,stripe}$ spectrum, with several of its peaks appearing in both $m_\mathrm{x,0}$ and $R$. The lowest frequency peak, $f_\mathrm{pseudo}$, is identified as the modulation frequency, just as it was for 10.4 mA. However, some of the peaks in the $m_\mathrm{z,0,stripe}$ spectrum are separated by smaller gaps (circled in black). This discrepancy means that $f_\mathrm{pseudo}$ is not a subharmonic (fraction) of $f_\mathrm{gyr}$, and as shown in Fig.~\ref{fig:OC_mod}(f), the magnetization dynamics are not perfectly periodic within the corresponding pseudo-period, $T_\mathrm{pseudo}$. Consequently, the orbits differ, as the pseudo-orbit (trajectory over $T_\mathrm{pseudo}$, shown in Fig.~\ref{fig:OC_mod}(g)) shifts slightly after every $T_\mathrm{pseudo}$, and the core does not appear to return to its initial position, leading to the position density plot in Fig.~\ref{fig:OC_mod}(h).

The two types of dynamics observed in this multi-frequency regime resemble  the commensurate and incommensurate states described in Ref.~\onlinecite{yoo_pattern_2020}. In the commensurate cases, such as at 10.4 mA, the dynamics are periodic and have a ratio $f_\mathrm{mod}=f_\mathrm{gyr}/n$, where $n$ is the number of full turns undergone by the vortex core to return to the same position along the orbit. In contrast, the incommensurate cases correspond to aperiodic (seemingly chaotic) regimes, such as at 12.8 mA, where the modulation frequency is not well defined and the orbits do not close. This type of dynamics arises from competition between two frequencies in the system, as exemplified in the Frenkel-Kontorowa model of commensurate-incommensurate phase transitions~\cite{bak_commensurate_1982, chaikin_principles_1995}. In our case, we suspect the two competing frequencies to be the gyration eigenfrequencies of each layer in the presence of the Oersted field, which differ from one another due to their chiralities. The evolution of these eigenfrequencies as a function of the current is shown for an uncoupled single layer in Fig.~\ref{fig:OC_mod}(i), for both positive (diamond markers) and negative (circle markers) chiralities. As the Oersted field (current) increases, the frequency for chirality parallel (anti-parallel) to the field, $c=-1$ ($c=+1$), increases (decreases). Consequently, the gap between eigenfrequencies widens with increasing current, up to the point where synchronization becomes weaker, and modulations become apparent.

The dynamics presented here for opposite chiralities only occur when the P-favored layer ($m_1$ here) has a chirality parallel to the Oersted field, while the AP-favored layer ($m_0$ here) is antiparallel to it. If the chiralities are inverted, the chirality opposed to the Oersted field switches at lower currents, resulting in C-state dynamics. The difference in dynamics between these two cases is easily understood through the frustration in each layer. In the case discussed here, both layers are partially frustrated. Since $m_1$ favors P configuration, having $c_0 = -c_1$ induces its frustration, while the frustration in $m_0$ is induced by the anti-alignment of its chirality to the Oersted field. In contrast, in the second case, $m_0$ is not frustrated, and $m_1$ is strongly frustrated, which reduces the current value required for the chirality to switch. This first study demonstrates that considering asymmetrical system configurations leads to interesting complex gyration regimes in the windmill system.

\section{Effect on asymmetry in $M_{s,i} d_i$ }\label{sec:asymmetry}

\begin{figure*}
\centering
	\includegraphics[width=17cm]{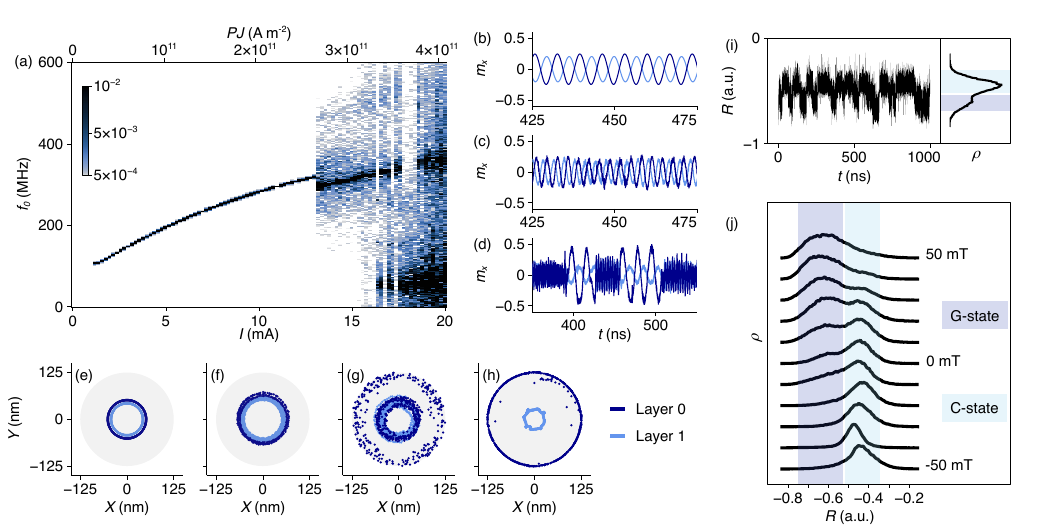}
	\caption{\label{fig7:asymmetrical Ms overview} Overview of dynamics for $p_0=p_1=1$, $c_0=c_1=-1$ and $M_{s,1} d_1 = 1.2M_{s,0} d_0$. (a) PSD map of the gyration frequency in layer 0 as a function of the current. (b), (c) and (d) are respectively the time traces of $m_x$ in each layer at currents $I=$3, 15 and 19 mA respectively. Light blue plots correspond to layer 1 and dark blue to layer 0. (e) and (f) are the trajectories of the vortex cores at currents $I=$3 and 15 mA. (g) and (h) are the trajectories of the vortex cores at current $I=$19 mA, for C- and G- states respectively. In (g), core centers appearing at the disk edges correspond to the C-state. (i) Time trace  of the magnetoresistance, $R = - \int  m_{0}(\mathbf{r}) \cdot m_{1}(\mathbf{r})\text{ } d\mathbf{r}^2 /\pi r^{2}_{disk}$, at current $I=$20 mA (left panel) and resistance distribution $\rho$ (right panel). Two peaks observed at R$\approx$-0.44 (light blue area, C-state) and R$\approx$-0.61 (dark blue area, G-state). (j) Resistance distribution at current $I=20$~mA and perpendicular field $B_z$ going from -50 to 50 mT. Negative (positive) fields favor the C-state (G-state).}
\end{figure*}

In Ref.~\onlinecite{locatelli_dynamics_2011}, it was demonstrated that the vortex polarity and chirality can be independently controlled for each layer, making the three studied configurations experimentally feasible. However, the relative vortex configurations are not the sole parameter influencing the dynamics. For instance, the geometry and magnetic material used for the free layers can also be varied, and would be easier to control. In this section, we investigate the impact of asymmetry in the geometry to assess its effects in a more experimentally viable system. To ensure that the observed dynamics stem solely from geometry asymmetry, we focus on the IV vortex configuration, with $p_0=p_1=1$ and $c_0=c_1=-1$. We consider an asymmetry in the $M_{s} d$ product, which models free layers with varying thicknesses. This asymmetry is controlled by the $\zeta$ ratio, such that $M_{s,1} d_1 = \zeta M_{s,0} d_0$. When $\zeta<1$, $\beta_1>\beta_0$ (Eq.~\ref{eq:betaSTT}), indicating that the STT in layer 1 will dominate the dynamics. In our current configuration (Fig.~\ref{fig:System}), layer 2 is favored for the P configuration, so $\zeta<1$ results in both layers overlapping, effectively eliminating the dynamics. Therefore, we choose $\zeta>1$, where $\beta_0>\beta_1$, ensuring that the STT favors an AP configuration and prevents the vortices from overlapping.

Figure~\ref{fig7:asymmetrical Ms overview} summarizes the dynamics obtained for $\zeta = 1.2$. Figure~\ref{fig7:asymmetrical Ms overview}(a) displays the power spectral density map as a function of the applied current, computed as explained in Sec.~\ref{sec:relativeConfiguration}, over the last $t_0 = 650$ ns of 1000 ns simulations. This map reveals three distinct dynamical regimes. Figures~\ref{fig7:asymmetrical Ms overview}(b), (c), and (d) present time traces of the magnetization component along the $x-$axis for both layers within these regimes. Figures~\ref{fig7:asymmetrical Ms overview}(e) to \ref{fig7:asymmetrical Ms overview}(h) illustrate the corresponding core trajectories for each regime.

We observe that the first dynamical regime commences at $I = 1$~mA, significantly lower than in the IV configuration of layers with equal thickness. It extends up to $I \approx 13$~mA. In this regime, both cores enter self-sustained gyration within their respective layers, transitioning into the G-state. The predicted in-plane magnetization signal exhibits distinct sinusoidal variations [Figure~\ref{fig7:asymmetrical Ms overview}(b)]. These variations arise from well-defined circular orbits of the vortex cores. In this stable regime, the two cores maintain phase opposition ($\delta \Phi = \pi$) at the lowest currents. However, as the current increases, the phase differences tend to $\delta \Phi = \pi/2$.

The current range $I = [13, 16]$~mA marks the transition into the second regime. The $m_x$ time traces and orbits illustrate G-state dynamics in both layers, with $\delta \Phi = \pi/2$ and fluctuations [Figs.~\ref{fig7:asymmetrical Ms overview}(c),(f)]. These fluctuations significantly broaden the power spectra. Notably, in the regime $I>16$~mA, a second frequency band emerges around $f_0=58$~MHz. This low-frequency band resembles those observed in the C-state dynamics previously discussed. However, its coexistence with the regular gyration band suggests more intricate dynamics. Figure \ref{fig7:asymmetrical Ms overview}(d) presents the $m_x$ time traces obtained in this regime. Distinct transitions between two oscillation regimes are evident. The high-frequency, low-amplitude regime corresponds to a fluctuating G-state, similar to the one depicted in Figure~\ref{fig7:asymmetrical Ms overview}(c). Conversely, the low-frequency, high-amplitude variations correspond to a windmill C-state, where the core in layer 0 is pushed outside the disk. The trajectories in Figs. \ref{fig7:asymmetrical Ms overview}(g) (C-state) and \ref{fig7:asymmetrical Ms overview}(h) (G-state) further substantiate this observation. These transitions appear to be chaotic, occurring through core reversal of vortex 0, after interactions with vortex/antivortex pairs generated due to the strong currents, as observed in the IV case with $\zeta=1$ (Fig.~\ref{fig:CGS transition}). While these chaotic transitions could also occur in the $[13,16]$~mA range, at these currents, fluctuations and vortex/anti-vortex creation are weaker, requiring longer simulations to observe them.

In the left panel of Fig.~\ref{fig7:asymmetrical Ms overview}(i), a time trace of the magnetoresistance at 20 mA [as computed with Eq.~(\ref{eq:MR})] is presented. This trace indicates that the minimum resistance state (parallel magnetic states) corresponds to a resistance of -1, while the maximum resistance state corresponds to a resistance of 1. The right panel displays the resistance distribution during this time trace. However, the resulting magnetoresistance does not accurately represent the gyration frequencies of the two different states due to fluctuations. Despite this, we observe two distinct resistance levels, each corresponding to one of the gyrotropic states. The lower resistance level, approximately $-0.61$, is attributed to the G-state, where the distance between cores is smaller. This results in a more parallel configuration since $c_0 = c_1$. The higher resistance level, approximately $-0.44$, corresponds to the C-state. Therefore, these chaotic transitions should be electrically measurable in experiments.

Figure~\ref{fig7:asymmetrical Ms overview}(j) presents the magnetoresistance distribution for different simulations at 20 mA, each lasting 500 ns, with an external perpendicular field applied. The applied fields range from -50 to 50 mA. When a positive field is applied, the distribution shifts towards the value of the G-state (dark blue area), while a negative field shifts it towards the C-state (light blue area). This demonstrates that the stability and probability of the two states can be tuned by an external field.

\section{Discussion and concluding remarks}

We have presented a detailed computational overview of the predicted dynamics in a double vortex windmill oscillator using our implementation of the coupled LLG equations. In all the results shown, each layer was discretized using a $256 \times 256 \times 1$ mesh, and the spacer layer was neglected. This allowed for a relatively small total mesh size ($256 \times 256 \times 2$) for more time-efficient computation. However, neglecting the spacer layer means that no physical separation between the two layers is simulated. This does not affect the interlayer exchange interaction, as it is set to zero, but it does make the distance-dependent dipole-dipole interaction between layers stronger.

To verify the validity of our results, we performed additional test simulations for each of the different regimes in IV, OP, and OC configurations with full discretization along the film thickness. We used a mesh of $128 \times 128 \times 11$, accounting for a non-magnetic spacer layer (layer 0, 7 nm: five sub-layers; spacer layer, 1.4 nm: one sub-layer; layer 1, 7 nm: five sub-layers). In these simulations, the STT only occurs at the interface between the layers (the sub-layers in contact with the spacer). The behaviors observed remain qualitatively unchanged, except for the modulated regime in the OC configuration. When the spacer layer is added, this regime becomes unstable and transitions into a more stable one where the core trajectories become elliptical.

The $256 \times 256 \times 2$ mesh also means that the two layers have the same thickness $d$, which is why in Sec.~\ref{sec:asymmetry} we model the asymmetry in thickness through $M_{s}$. This approach also leads to some corrections in the dipole-dipole term. Increasing $M_{s}$ will increase the density of magnetic moments inside the layer and not its size, leading to a smaller distance between the new moments and the second layer and a slightly stronger magnetostatic field. We also performed further test simulations for $I =$ 10 and 20 mA shown in Sec.~\ref{sec:asymmetry}, using a mesh of $256 \times 256 \times 12$ that truly simulates asymmetric thicknesses and a spacer layer [layer 0 (8.4 nm): six sub-layers, spacer layer (1.4 nm): one sub-layer, layer 1 (7 nm): five sub-layers]. As expected, we observe regular gyration at 10 mA, and chaotic transitions at 20 mA. A difference in the transition rate is observed, however due to the chaotic nature it is difficult to obtain useful metrics for the comparison.

In our model, the current density is assumed to be uniform throughout the entire disk, disregarding any dependence on the scalar product $m_0 \cdot m_1$ that governs the magnetoresistance. For coupled vortices, the primary STT interaction occurs between one layer’s core and the other layer’s in-plane magnetization due to the cross product $m_0 \times m_1 $ (Eq.~\ref{eq:LLGS}). Consequently, the magnetoresistance, which is proportional to $m_0 \cdot m_1$, would have a nearly identical value in all cells contributing to the STT. Therefore, our uniform current density approximation should provide a realistic description of the dynamics, particularly regarding the resistance contribution to the STT. However, it is crucial to note that for a magnetic tunnel junction (MTJ), one would also need to consider the dynamical changes in local current densities resulting from the transport properties of the MTJ, which we do not account for in this model. This omission could lead to discrepancies between predicted and experimental behavior.

A common feature of the complex dynamics observed is the transition to the \emph{windmill C-state}. In this state, the AP- favoring layer transitions from regular gyration, while the vortex in the P-favoring layer maintains it. This \emph{windmill C-state} differs from the C-state in regular STVOs (with a single free layer) because the gyration frequency decreases with current instead of increasing. This difference in behavior could be attributed to several reasons, such as the absence of perpendicular field in our system compared to the very strong fields used to observe experimental C-states (as discussed in Ref.~\onlinecite{wittrock_beyond_2021}).

In summary, we have provided a detailed account of the dynamics in a spin-torque oscillator with two vortex-free layers. Our simulations show how asymmetry between these layers is crucial. This asymmetry can be introduced either by differences in chirality and polarity between the vortex configurations or by variations in thickness and material. This asymmetry allows us to observe a dynamic C-state in the Slonczewski windmill, a phenomenon that does not require external fields, unlike single free-layer oscillators. These oscillators exhibit a range of dynamical behaviors depending on the applied currents. We also predict that the chaotic transitions between the C-state and regular gyration states, mediated by core reversals in one of the layers, are measurable and tunable. These transitions could potentially be used to generate random numbers for encryption and probabilistic computing, while complex transients could be useful for pattern recognition tasks.

\begin{acknowledgments}
We thank Ursula Ebels and Thibaut Devolder for fruitful discussions. This work was supported by the French government under the France 2030 investment plan, PEPR SPIN, managed by the Agence Nationale de la Recherche under contract number ANR-22-EXSP-0005 (SPINCOM).
\end{acknowledgments}

\bibliography{windmill}

\begin{thebibliography}{37}%
\makeatletter
\providecommand \@ifxundefined [1]{%
 \@ifx{#1\undefined}
}%
\providecommand \@ifnum [1]{%
 \ifnum #1\expandafter \@firstoftwo
 \else \expandafter \@secondoftwo
 \fi
}%
\providecommand \@ifx [1]{%
 \ifx #1\expandafter \@firstoftwo
 \else \expandafter \@secondoftwo
 \fi
}%
\providecommand \natexlab [1]{#1}%
\providecommand \enquote  [1]{``#1''}%
\providecommand \bibnamefont  [1]{#1}%
\providecommand \bibfnamefont [1]{#1}%
\providecommand \citenamefont [1]{#1}%
\providecommand \href@noop [0]{\@secondoftwo}%
\providecommand \href [0]{\begingroup \@sanitize@url \@href}%
\providecommand \@href[1]{\@@startlink{#1}\@@href}%
\providecommand \@@href[1]{\endgroup#1\@@endlink}%
\providecommand \@sanitize@url [0]{\catcode `\\12\catcode `\$12\catcode `\&12\catcode `\#12\catcode `\^12\catcode `\_12\catcode `\%12\relax}%
\providecommand \@@startlink[1]{}%
\providecommand \@@endlink[0]{}%
\providecommand \url  [0]{\begingroup\@sanitize@url \@url }%
\providecommand \@url [1]{\endgroup\@href {#1}{\urlprefix }}%
\providecommand \urlprefix  [0]{URL }%
\providecommand \Eprint [0]{\href }%
\providecommand \doibase [0]{https://doi.org/}%
\providecommand \selectlanguage [0]{\@gobble}%
\providecommand \bibinfo  [0]{\@secondoftwo}%
\providecommand \bibfield  [0]{\@secondoftwo}%
\providecommand \translation [1]{[#1]}%
\providecommand \BibitemOpen [0]{}%
\providecommand \bibitemStop [0]{}%
\providecommand \bibitemNoStop [0]{.\EOS\space}%
\providecommand \EOS [0]{\spacefactor3000\relax}%
\providecommand \BibitemShut  [1]{\csname bibitem#1\endcsname}%
\let\auto@bib@innerbib\@empty
\bibitem [{\citenamefont {Slonczewski}(1996)}]{slonczewski_current-driven_1996}%
  \BibitemOpen
  \bibfield  {author} {\bibinfo {author} {\bibfnamefont {J.}~\bibnamefont {Slonczewski}},\ }\bibfield  {title} {\bibinfo {title} {Current-driven excitation of magnetic multilayers},\ }\href {https://doi.org/10.1016/0304-8853(96)00062-5} {\bibfield  {journal} {\bibinfo  {journal} {Journal of Magnetism and Magnetic Materials}\ }\textbf {\bibinfo {volume} {159}},\ \bibinfo {pages} {L1} (\bibinfo {year} {1996})}\BibitemShut {NoStop}%
\bibitem [{\citenamefont {Berger}(1996)}]{berger_emission_1996}%
  \BibitemOpen
  \bibfield  {author} {\bibinfo {author} {\bibfnamefont {L.}~\bibnamefont {Berger}},\ }\bibfield  {title} {\bibinfo {title} {{Emission of spin waves by a magnetic multilayer traversed by a current}},\ }\href {https://doi.org/10.1103/physrevb.54.9353} {\bibfield  {journal} {\bibinfo  {journal} {Physical Review B}\ }\textbf {\bibinfo {volume} {54}},\ \bibinfo {pages} {9353 } (\bibinfo {year} {1996})}\BibitemShut {NoStop}%
\bibitem [{\citenamefont {Kiselev}\ \emph {et~al.}(2003)\citenamefont {Kiselev}, \citenamefont {Sankey}, \citenamefont {Krivorotov}, \citenamefont {Emley}, \citenamefont {Schoelkopf}, \citenamefont {Buhrman},\ and\ \citenamefont {Ralph}}]{kiselev_microwave_2003}%
  \BibitemOpen
  \bibfield  {author} {\bibinfo {author} {\bibfnamefont {S.~I.}\ \bibnamefont {Kiselev}}, \bibinfo {author} {\bibfnamefont {J.~C.}\ \bibnamefont {Sankey}}, \bibinfo {author} {\bibfnamefont {I.~N.}\ \bibnamefont {Krivorotov}}, \bibinfo {author} {\bibfnamefont {N.~C.}\ \bibnamefont {Emley}}, \bibinfo {author} {\bibfnamefont {R.~J.}\ \bibnamefont {Schoelkopf}}, \bibinfo {author} {\bibfnamefont {R.~A.}\ \bibnamefont {Buhrman}},\ and\ \bibinfo {author} {\bibfnamefont {D.~C.}\ \bibnamefont {Ralph}},\ }\bibfield  {title} {\bibinfo {title} {Microwave oscillations of a nanomagnet driven by a spin-polarized current},\ }\href {https://doi.org/10.1038/nature01967} {\bibfield  {journal} {\bibinfo  {journal} {Nature}\ }\textbf {\bibinfo {volume} {425}},\ \bibinfo {pages} {380} (\bibinfo {year} {2003})}\BibitemShut {NoStop}%
\bibitem [{\citenamefont {Rippard}\ \emph {et~al.}(2004)\citenamefont {Rippard}, \citenamefont {Pufall}, \citenamefont {Kaka}, \citenamefont {Russek},\ and\ \citenamefont {Silva}}]{rippard_direct_2004}%
  \BibitemOpen
  \bibfield  {author} {\bibinfo {author} {\bibfnamefont {W.}~\bibnamefont {Rippard}}, \bibinfo {author} {\bibfnamefont {M.}~\bibnamefont {Pufall}}, \bibinfo {author} {\bibfnamefont {S.}~\bibnamefont {Kaka}}, \bibinfo {author} {\bibfnamefont {S.}~\bibnamefont {Russek}},\ and\ \bibinfo {author} {\bibfnamefont {T.}~\bibnamefont {Silva}},\ }\bibfield  {title} {\bibinfo {title} {{Direct-Current Induced Dynamics in Co90Fe10/Ni80Fe20 Point Contacts}},\ }\href {https://doi.org/10.1103/physrevlett.92.027201} {\bibfield  {journal} {\bibinfo  {journal} {Physical Review Letters}\ }\textbf {\bibinfo {volume} {92}},\ \bibinfo {pages} {027201} (\bibinfo {year} {2004})}\BibitemShut {NoStop}%
\bibitem [{\citenamefont {Pribiag}\ \emph {et~al.}(2007)\citenamefont {Pribiag}, \citenamefont {Krivorotov}, \citenamefont {Fuchs}, \citenamefont {Braganca}, \citenamefont {Ozatay}, \citenamefont {Sankey}, \citenamefont {Ralph},\ and\ \citenamefont {Buhrman}}]{pribiag_magnetic_2007}%
  \BibitemOpen
  \bibfield  {author} {\bibinfo {author} {\bibfnamefont {V.~S.}\ \bibnamefont {Pribiag}}, \bibinfo {author} {\bibfnamefont {I.~N.}\ \bibnamefont {Krivorotov}}, \bibinfo {author} {\bibfnamefont {G.~D.}\ \bibnamefont {Fuchs}}, \bibinfo {author} {\bibfnamefont {P.~M.}\ \bibnamefont {Braganca}}, \bibinfo {author} {\bibfnamefont {O.}~\bibnamefont {Ozatay}}, \bibinfo {author} {\bibfnamefont {J.~C.}\ \bibnamefont {Sankey}}, \bibinfo {author} {\bibfnamefont {D.~C.}\ \bibnamefont {Ralph}},\ and\ \bibinfo {author} {\bibfnamefont {R.~A.}\ \bibnamefont {Buhrman}},\ }\bibfield  {title} {\bibinfo {title} {Magnetic vortex oscillator driven by d.c. spin-polarized current},\ }\href {https://doi.org/10.1038/nphys619} {\bibfield  {journal} {\bibinfo  {journal} {Nature Physics}\ }\textbf {\bibinfo {volume} {3}},\ \bibinfo {pages} {498} (\bibinfo {year} {2007})}\BibitemShut {NoStop}%
\bibitem [{\citenamefont {Mistral}\ \emph {et~al.}(2008)\citenamefont {Mistral}, \citenamefont {van Kampen}, \citenamefont {Hrkac}, \citenamefont {Kim}, \citenamefont {Devolder}, \citenamefont {Crozat}, \citenamefont {Chappert}, \citenamefont {Lagae},\ and\ \citenamefont {Schrefl}}]{mistral_current_2008}%
  \BibitemOpen
  \bibfield  {author} {\bibinfo {author} {\bibfnamefont {Q.}~\bibnamefont {Mistral}}, \bibinfo {author} {\bibfnamefont {M.}~\bibnamefont {van Kampen}}, \bibinfo {author} {\bibfnamefont {G.}~\bibnamefont {Hrkac}}, \bibinfo {author} {\bibfnamefont {J.-V.}\ \bibnamefont {Kim}}, \bibinfo {author} {\bibfnamefont {T.}~\bibnamefont {Devolder}}, \bibinfo {author} {\bibfnamefont {P.}~\bibnamefont {Crozat}}, \bibinfo {author} {\bibfnamefont {C.}~\bibnamefont {Chappert}}, \bibinfo {author} {\bibfnamefont {L.}~\bibnamefont {Lagae}},\ and\ \bibinfo {author} {\bibfnamefont {T.}~\bibnamefont {Schrefl}},\ }\bibfield  {title} {\bibinfo {title} {{Current-Driven Vortex Oscillations in Metallic Nanocontacts}},\ }\href {https://doi.org/10.1103/physrevlett.100.257201} {\bibfield  {journal} {\bibinfo  {journal} {Physical Review Letters}\ }\textbf {\bibinfo {volume} {100}},\ \bibinfo {pages} {257201} (\bibinfo {year} {2008})}\BibitemShut {NoStop}%
\bibitem [{\citenamefont {Garcia-Sanchez}\ \emph {et~al.}(2016)\citenamefont {Garcia-Sanchez}, \citenamefont {Sampaio}, \citenamefont {Reyren}, \citenamefont {Cros},\ and\ \citenamefont {Kim}}]{garciasanchez_skyrmion_2016}%
  \BibitemOpen
  \bibfield  {author} {\bibinfo {author} {\bibfnamefont {F.}~\bibnamefont {Garcia-Sanchez}}, \bibinfo {author} {\bibfnamefont {J.}~\bibnamefont {Sampaio}}, \bibinfo {author} {\bibfnamefont {N.}~\bibnamefont {Reyren}}, \bibinfo {author} {\bibfnamefont {V.}~\bibnamefont {Cros}},\ and\ \bibinfo {author} {\bibfnamefont {J.-V.}\ \bibnamefont {Kim}},\ }\bibfield  {title} {\bibinfo {title} {{A skyrmion-based spin-torque nano-oscillator}},\ }\href {https://doi.org/10.1088/1367-2630/18/7/075011} {\bibfield  {journal} {\bibinfo  {journal} {New Journal of Physics}\ }\textbf {\bibinfo {volume} {18}},\ \bibinfo {pages} {075011} (\bibinfo {year} {2016})}\BibitemShut {NoStop}%
\bibitem [{\citenamefont {Wittrock}\ \emph {et~al.}(2021)\citenamefont {Wittrock}, \citenamefont {Talatchian}, \citenamefont {Romera}, \citenamefont {Menshawy}, \citenamefont {Jotta~Garcia}, \citenamefont {Cyrille}, \citenamefont {Ferreira}, \citenamefont {Lebrun}, \citenamefont {Bortolotti}, \citenamefont {Ebels}, \citenamefont {Grollier},\ and\ \citenamefont {Cros}}]{wittrock_beyond_2021}%
  \BibitemOpen
  \bibfield  {author} {\bibinfo {author} {\bibfnamefont {S.}~\bibnamefont {Wittrock}}, \bibinfo {author} {\bibfnamefont {P.}~\bibnamefont {Talatchian}}, \bibinfo {author} {\bibfnamefont {M.}~\bibnamefont {Romera}}, \bibinfo {author} {\bibfnamefont {S.}~\bibnamefont {Menshawy}}, \bibinfo {author} {\bibfnamefont {M.}~\bibnamefont {Jotta~Garcia}}, \bibinfo {author} {\bibfnamefont {M.-C.}\ \bibnamefont {Cyrille}}, \bibinfo {author} {\bibfnamefont {R.}~\bibnamefont {Ferreira}}, \bibinfo {author} {\bibfnamefont {R.}~\bibnamefont {Lebrun}}, \bibinfo {author} {\bibfnamefont {P.}~\bibnamefont {Bortolotti}}, \bibinfo {author} {\bibfnamefont {U.}~\bibnamefont {Ebels}}, \bibinfo {author} {\bibfnamefont {J.}~\bibnamefont {Grollier}},\ and\ \bibinfo {author} {\bibfnamefont {V.}~\bibnamefont {Cros}},\ }\bibfield  {title} {\bibinfo {title} {Beyond the gyrotropic motion: {Dynamic} {C}-state in vortex spin torque oscillators},\ }\href {https://doi.org/10.1063/5.0029083} {\bibfield  {journal} {\bibinfo  {journal} {Applied
  Physics Letters}\ }\textbf {\bibinfo {volume} {118}},\ \bibinfo {pages} {012404} (\bibinfo {year} {2021})}\BibitemShut {NoStop}%
\bibitem [{\citenamefont {Villard}\ \emph {et~al.}(2010)\citenamefont {Villard}, \citenamefont {Ebels}, \citenamefont {Houssameddine}, \citenamefont {Katine}, \citenamefont {Mauri}, \citenamefont {Delaet}, \citenamefont {Vincent}, \citenamefont {Cyrille}, \citenamefont {Viala}, \citenamefont {Michel}, \citenamefont {Prouvee},\ and\ \citenamefont {Badets}}]{villard_ghz_2010}%
  \BibitemOpen
  \bibfield  {author} {\bibinfo {author} {\bibfnamefont {P.}~\bibnamefont {Villard}}, \bibinfo {author} {\bibfnamefont {U.}~\bibnamefont {Ebels}}, \bibinfo {author} {\bibfnamefont {D.}~\bibnamefont {Houssameddine}}, \bibinfo {author} {\bibfnamefont {J.}~\bibnamefont {Katine}}, \bibinfo {author} {\bibfnamefont {D.}~\bibnamefont {Mauri}}, \bibinfo {author} {\bibfnamefont {B.}~\bibnamefont {Delaet}}, \bibinfo {author} {\bibfnamefont {P.}~\bibnamefont {Vincent}}, \bibinfo {author} {\bibfnamefont {M.-C.}\ \bibnamefont {Cyrille}}, \bibinfo {author} {\bibfnamefont {B.}~\bibnamefont {Viala}}, \bibinfo {author} {\bibfnamefont {J.-P.}\ \bibnamefont {Michel}}, \bibinfo {author} {\bibfnamefont {J.}~\bibnamefont {Prouvee}},\ and\ \bibinfo {author} {\bibfnamefont {F.}~\bibnamefont {Badets}},\ }\bibfield  {title} {\bibinfo {title} {{A GHz Spintronic-Based RF Oscillator}},\ }\href {https://doi.org/10.1109/jssc.2009.2034432} {\bibfield  {journal} {\bibinfo  {journal} {IEEE Journal of Solid-State Circuits}\ }\textbf {\bibinfo
  {volume} {45}},\ \bibinfo {pages} {214 } (\bibinfo {year} {2010})}\BibitemShut {NoStop}%
\bibitem [{\citenamefont {Litvinenko}\ \emph {et~al.}(2021)\citenamefont {Litvinenko}, \citenamefont {Sethi}, \citenamefont {Murapaka}, \citenamefont {Jenkins}, \citenamefont {Cros}, \citenamefont {Bortolotti}, \citenamefont {Ferreira}, \citenamefont {Dieny},\ and\ \citenamefont {Ebels}}]{litvinenko_analog_2021}%
  \BibitemOpen
  \bibfield  {author} {\bibinfo {author} {\bibfnamefont {A.}~\bibnamefont {Litvinenko}}, \bibinfo {author} {\bibfnamefont {P.}~\bibnamefont {Sethi}}, \bibinfo {author} {\bibfnamefont {C.}~\bibnamefont {Murapaka}}, \bibinfo {author} {\bibfnamefont {A.}~\bibnamefont {Jenkins}}, \bibinfo {author} {\bibfnamefont {V.}~\bibnamefont {Cros}}, \bibinfo {author} {\bibfnamefont {P.}~\bibnamefont {Bortolotti}}, \bibinfo {author} {\bibfnamefont {R.}~\bibnamefont {Ferreira}}, \bibinfo {author} {\bibfnamefont {B.}~\bibnamefont {Dieny}},\ and\ \bibinfo {author} {\bibfnamefont {U.}~\bibnamefont {Ebels}},\ }\bibfield  {title} {\bibinfo {title} {{Analog and Digital Phase Modulation and Signal Transmission with Spin-Torque Nano-Oscillators}},\ }\href {https://doi.org/10.1103/physrevapplied.16.024048} {\bibfield  {journal} {\bibinfo  {journal} {Physical Review Applied}\ }\textbf {\bibinfo {volume} {16}},\ \bibinfo {pages} {024048} (\bibinfo {year} {2021})},\ \Eprint {https://arxiv.org/abs/1905.02443} {1905.02443} \BibitemShut
  {NoStop}%
\bibitem [{\citenamefont {Zhu}\ \emph {et~al.}(2008)\citenamefont {Zhu}, \citenamefont {Zhu},\ and\ \citenamefont {Tang}}]{zhu_microwave_2008}%
  \BibitemOpen
  \bibfield  {author} {\bibinfo {author} {\bibfnamefont {J.-G.}\ \bibnamefont {Zhu}}, \bibinfo {author} {\bibfnamefont {X.}~\bibnamefont {Zhu}},\ and\ \bibinfo {author} {\bibfnamefont {Y.}~\bibnamefont {Tang}},\ }\bibfield  {title} {\bibinfo {title} {{Microwave Assisted Magnetic Recording}},\ }\href {https://doi.org/10.1109/tmag.2007.911031} {\bibfield  {journal} {\bibinfo  {journal} {IEEE Transactions on Magnetics}\ }\textbf {\bibinfo {volume} {44}},\ \bibinfo {pages} {125} (\bibinfo {year} {2008})}\BibitemShut {NoStop}%
\bibitem [{\citenamefont {Zhu}\ and\ \citenamefont {Wang}(2010)}]{zhu_microwave_2010}%
  \BibitemOpen
  \bibfield  {author} {\bibinfo {author} {\bibfnamefont {J.-G.}\ \bibnamefont {Zhu}}\ and\ \bibinfo {author} {\bibfnamefont {Y.}~\bibnamefont {Wang}},\ }\bibfield  {title} {\bibinfo {title} {{Microwave Assisted Magnetic Recording Utilizing Perpendicular Spin Torque Oscillator With Switchable Perpendicular Electrodes}},\ }\href {https://doi.org/10.1109/tmag.2009.2036588} {\bibfield  {journal} {\bibinfo  {journal} {IEEE Transactions on Magnetics}\ }\textbf {\bibinfo {volume} {46}},\ \bibinfo {pages} {751} (\bibinfo {year} {2010})}\BibitemShut {NoStop}%
\bibitem [{\citenamefont {Torrejon}\ \emph {et~al.}(2017)\citenamefont {Torrejon}, \citenamefont {Riou}, \citenamefont {Araujo}, \citenamefont {Tsunegi}, \citenamefont {Khalsa}, \citenamefont {Querlioz}, \citenamefont {Bortolotti}, \citenamefont {Cros}, \citenamefont {Yakushiji}, \citenamefont {Fukushima}, \citenamefont {Kubota}, \citenamefont {Yuasa}, \citenamefont {Stiles},\ and\ \citenamefont {Grollier}}]{torrejon_neuromorphic_2017}%
  \BibitemOpen
  \bibfield  {author} {\bibinfo {author} {\bibfnamefont {J.}~\bibnamefont {Torrejon}}, \bibinfo {author} {\bibfnamefont {M.}~\bibnamefont {Riou}}, \bibinfo {author} {\bibfnamefont {F.~A.}\ \bibnamefont {Araujo}}, \bibinfo {author} {\bibfnamefont {S.}~\bibnamefont {Tsunegi}}, \bibinfo {author} {\bibfnamefont {G.}~\bibnamefont {Khalsa}}, \bibinfo {author} {\bibfnamefont {D.}~\bibnamefont {Querlioz}}, \bibinfo {author} {\bibfnamefont {P.}~\bibnamefont {Bortolotti}}, \bibinfo {author} {\bibfnamefont {V.}~\bibnamefont {Cros}}, \bibinfo {author} {\bibfnamefont {K.}~\bibnamefont {Yakushiji}}, \bibinfo {author} {\bibfnamefont {A.}~\bibnamefont {Fukushima}}, \bibinfo {author} {\bibfnamefont {H.}~\bibnamefont {Kubota}}, \bibinfo {author} {\bibfnamefont {S.}~\bibnamefont {Yuasa}}, \bibinfo {author} {\bibfnamefont {M.~D.}\ \bibnamefont {Stiles}},\ and\ \bibinfo {author} {\bibfnamefont {J.}~\bibnamefont {Grollier}},\ }\bibfield  {title} {\bibinfo {title} {{Neuromorphic computing with nanoscale spintronic oscillators}},\
  }\href {https://doi.org/10.1038/nature23011} {\bibfield  {journal} {\bibinfo  {journal} {Nature}\ }\textbf {\bibinfo {volume} {547}},\ \bibinfo {pages} {428 } (\bibinfo {year} {2017})}\BibitemShut {NoStop}%
\bibitem [{\citenamefont {Romera}\ \emph {et~al.}(2018)\citenamefont {Romera}, \citenamefont {Talatchian}, \citenamefont {Tsunegi}, \citenamefont {Araujo}, \citenamefont {Cros}, \citenamefont {Bortolotti}, \citenamefont {Trastoy}, \citenamefont {Yakushiji}, \citenamefont {Fukushima}, \citenamefont {Kubota}, \citenamefont {Yuasa}, \citenamefont {Ernoult}, \citenamefont {Vodenicarevic}, \citenamefont {Hirtzlin}, \citenamefont {Locatelli}, \citenamefont {Querlioz},\ and\ \citenamefont {Grollier}}]{romera_vowel_2018}%
  \BibitemOpen
  \bibfield  {author} {\bibinfo {author} {\bibfnamefont {M.}~\bibnamefont {Romera}}, \bibinfo {author} {\bibfnamefont {P.}~\bibnamefont {Talatchian}}, \bibinfo {author} {\bibfnamefont {S.}~\bibnamefont {Tsunegi}}, \bibinfo {author} {\bibfnamefont {F.~A.}\ \bibnamefont {Araujo}}, \bibinfo {author} {\bibfnamefont {V.}~\bibnamefont {Cros}}, \bibinfo {author} {\bibfnamefont {P.}~\bibnamefont {Bortolotti}}, \bibinfo {author} {\bibfnamefont {J.}~\bibnamefont {Trastoy}}, \bibinfo {author} {\bibfnamefont {K.}~\bibnamefont {Yakushiji}}, \bibinfo {author} {\bibfnamefont {A.}~\bibnamefont {Fukushima}}, \bibinfo {author} {\bibfnamefont {H.}~\bibnamefont {Kubota}}, \bibinfo {author} {\bibfnamefont {S.}~\bibnamefont {Yuasa}}, \bibinfo {author} {\bibfnamefont {M.}~\bibnamefont {Ernoult}}, \bibinfo {author} {\bibfnamefont {D.}~\bibnamefont {Vodenicarevic}}, \bibinfo {author} {\bibfnamefont {T.}~\bibnamefont {Hirtzlin}}, \bibinfo {author} {\bibfnamefont {N.}~\bibnamefont {Locatelli}}, \bibinfo {author} {\bibfnamefont
  {D.}~\bibnamefont {Querlioz}},\ and\ \bibinfo {author} {\bibfnamefont {J.}~\bibnamefont {Grollier}},\ }\bibfield  {title} {\bibinfo {title} {{Vowel recognition with four coupled spin-torque nano-oscillators}},\ }\href {https://doi.org/10.1038/s41586-018-0632-y} {\bibfield  {journal} {\bibinfo  {journal} {Nature}\ }\textbf {\bibinfo {volume} {563}},\ \bibinfo {pages} {230 } (\bibinfo {year} {2018})}\BibitemShut {NoStop}%
\bibitem [{\citenamefont {Riou}\ \emph {et~al.}(2019)\citenamefont {Riou}, \citenamefont {Torrejon}, \citenamefont {Garitaine}, \citenamefont {Abreu~Araujo}, \citenamefont {Bortolotti}, \citenamefont {Cros}, \citenamefont {Tsunegi}, \citenamefont {Yakushiji}, \citenamefont {Fukushima}, \citenamefont {Kubota}, \citenamefont {Yuasa}, \citenamefont {Querlioz}, \citenamefont {Stiles},\ and\ \citenamefont {Grollier}}]{riou_temporal_2019}%
  \BibitemOpen
  \bibfield  {author} {\bibinfo {author} {\bibfnamefont {M.}~\bibnamefont {Riou}}, \bibinfo {author} {\bibfnamefont {J.}~\bibnamefont {Torrejon}}, \bibinfo {author} {\bibfnamefont {B.}~\bibnamefont {Garitaine}}, \bibinfo {author} {\bibfnamefont {F.}~\bibnamefont {Abreu~Araujo}}, \bibinfo {author} {\bibfnamefont {P.}~\bibnamefont {Bortolotti}}, \bibinfo {author} {\bibfnamefont {V.}~\bibnamefont {Cros}}, \bibinfo {author} {\bibfnamefont {S.}~\bibnamefont {Tsunegi}}, \bibinfo {author} {\bibfnamefont {K.}~\bibnamefont {Yakushiji}}, \bibinfo {author} {\bibfnamefont {A.}~\bibnamefont {Fukushima}}, \bibinfo {author} {\bibfnamefont {H.}~\bibnamefont {Kubota}}, \bibinfo {author} {\bibfnamefont {S.}~\bibnamefont {Yuasa}}, \bibinfo {author} {\bibfnamefont {D.}~\bibnamefont {Querlioz}}, \bibinfo {author} {\bibfnamefont {M.}~\bibnamefont {Stiles}},\ and\ \bibinfo {author} {\bibfnamefont {J.}~\bibnamefont {Grollier}},\ }\bibfield  {title} {\bibinfo {title} {Temporal {Pattern} {Recognition} with {Delayed}-{Feedback}
  {Spin}-{Torque} {Nano}-{Oscillators}},\ }\href {https://doi.org/10.1103/PhysRevApplied.12.024049} {\bibfield  {journal} {\bibinfo  {journal} {Physical Review Applied}\ }\textbf {\bibinfo {volume} {12}},\ \bibinfo {pages} {024049} (\bibinfo {year} {2019})}\BibitemShut {NoStop}%
\bibitem [{\citenamefont {Imai}\ and\ \citenamefont {Taniguchi}(2023)}]{imai_associative_2023}%
  \BibitemOpen
  \bibfield  {author} {\bibinfo {author} {\bibfnamefont {Y.}~\bibnamefont {Imai}}\ and\ \bibinfo {author} {\bibfnamefont {T.}~\bibnamefont {Taniguchi}},\ }\bibfield  {title} {\bibinfo {title} {{Associative memory by virtual oscillator network based on single spin-torque oscillator}},\ }\href {https://doi.org/10.1038/s41598-023-42951-z} {\bibfield  {journal} {\bibinfo  {journal} {Scientific Reports}\ }\textbf {\bibinfo {volume} {13}},\ \bibinfo {pages} {15809} (\bibinfo {year} {2023})}\BibitemShut {NoStop}%
\bibitem [{\citenamefont {Phan}\ \emph {et~al.}(2024)\citenamefont {Phan}, \citenamefont {Prasad}, \citenamefont {Hakam}, \citenamefont {Valli}, \citenamefont {Anghel}, \citenamefont {Benetti}, \citenamefont {Madhavan}, \citenamefont {Jenkins}, \citenamefont {Ferreira}, \citenamefont {Stiles}, \citenamefont {Ebels},\ and\ \citenamefont {Talatchian}}]{phan_unbiased_2024}%
  \BibitemOpen
  \bibfield  {author} {\bibinfo {author} {\bibfnamefont {N.-T.}\ \bibnamefont {Phan}}, \bibinfo {author} {\bibfnamefont {N.}~\bibnamefont {Prasad}}, \bibinfo {author} {\bibfnamefont {A.}~\bibnamefont {Hakam}}, \bibinfo {author} {\bibfnamefont {A.~S.~E.}\ \bibnamefont {Valli}}, \bibinfo {author} {\bibfnamefont {L.}~\bibnamefont {Anghel}}, \bibinfo {author} {\bibfnamefont {L.}~\bibnamefont {Benetti}}, \bibinfo {author} {\bibfnamefont {A.}~\bibnamefont {Madhavan}}, \bibinfo {author} {\bibfnamefont {A.~S.}\ \bibnamefont {Jenkins}}, \bibinfo {author} {\bibfnamefont {R.}~\bibnamefont {Ferreira}}, \bibinfo {author} {\bibfnamefont {M.~D.}\ \bibnamefont {Stiles}}, \bibinfo {author} {\bibfnamefont {U.}~\bibnamefont {Ebels}},\ and\ \bibinfo {author} {\bibfnamefont {P.}~\bibnamefont {Talatchian}},\ }\bibfield  {title} {\bibinfo {title} {{Unbiased random bitstream generation using injection-locked spin-torque nano-oscillators}},\ }\href {https://doi.org/10.1103/physrevapplied.21.034063} {\bibfield  {journal} {\bibinfo
  {journal} {Physical Review Applied}\ }\textbf {\bibinfo {volume} {21}},\ \bibinfo {pages} {034063} (\bibinfo {year} {2024})}\BibitemShut {NoStop}%
\bibitem [{\citenamefont {Yoo}\ \emph {et~al.}(2020)\citenamefont {Yoo}, \citenamefont {Rontani}, \citenamefont {L{\'e}tang}, \citenamefont {Petit-Watelot}, \citenamefont {Devolder}, \citenamefont {Sciamanna}, \citenamefont {Bouzehouane}, \citenamefont {Cros},\ and\ \citenamefont {Kim}}]{yoo_pattern_2020}%
  \BibitemOpen
  \bibfield  {author} {\bibinfo {author} {\bibfnamefont {M.-W.}\ \bibnamefont {Yoo}}, \bibinfo {author} {\bibfnamefont {D.}~\bibnamefont {Rontani}}, \bibinfo {author} {\bibfnamefont {J.}~\bibnamefont {L{\'e}tang}}, \bibinfo {author} {\bibfnamefont {S.}~\bibnamefont {Petit-Watelot}}, \bibinfo {author} {\bibfnamefont {T.}~\bibnamefont {Devolder}}, \bibinfo {author} {\bibfnamefont {M.}~\bibnamefont {Sciamanna}}, \bibinfo {author} {\bibfnamefont {K.}~\bibnamefont {Bouzehouane}}, \bibinfo {author} {\bibfnamefont {V.}~\bibnamefont {Cros}},\ and\ \bibinfo {author} {\bibfnamefont {J.-V.}\ \bibnamefont {Kim}},\ }\bibfield  {title} {\bibinfo {title} {Pattern generation and symbolic dynamics in a nanocontact vortex oscillator},\ }\href {https://doi.org/10.1038/s41467-020-14328-7} {\bibfield  {journal} {\bibinfo  {journal} {Nature Communications}\ }\textbf {\bibinfo {volume} {11}},\ \bibinfo {pages} {601} (\bibinfo {year} {2020})}\BibitemShut {NoStop}%
\bibitem [{\citenamefont {Khvalkovskiy}\ \emph {et~al.}(2009)\citenamefont {Khvalkovskiy}, \citenamefont {Grollier}, \citenamefont {Dussaux}, \citenamefont {Zvezdin},\ and\ \citenamefont {Cros}}]{khvalkovskiy_vortex_2009}%
  \BibitemOpen
  \bibfield  {author} {\bibinfo {author} {\bibfnamefont {A.~V.}\ \bibnamefont {Khvalkovskiy}}, \bibinfo {author} {\bibfnamefont {J.}~\bibnamefont {Grollier}}, \bibinfo {author} {\bibfnamefont {A.}~\bibnamefont {Dussaux}}, \bibinfo {author} {\bibfnamefont {K.~A.}\ \bibnamefont {Zvezdin}},\ and\ \bibinfo {author} {\bibfnamefont {V.}~\bibnamefont {Cros}},\ }\bibfield  {title} {\bibinfo {title} {Vortex oscillations induced by spin-polarized current in a magnetic nanopillar: {Analytical} versus micromagnetic calculations},\ }\href {https://doi.org/10.1103/PhysRevB.80.140401} {\bibfield  {journal} {\bibinfo  {journal} {Physical Review B}\ }\textbf {\bibinfo {volume} {80}},\ \bibinfo {pages} {140401} (\bibinfo {year} {2009})}\BibitemShut {NoStop}%
\bibitem [{\citenamefont {Dussaux}\ \emph {et~al.}(2010)\citenamefont {Dussaux}, \citenamefont {Georges}, \citenamefont {Grollier}, \citenamefont {Cros}, \citenamefont {Khvalkovskiy}, \citenamefont {Fukushima}, \citenamefont {Konoto}, \citenamefont {Kubota}, \citenamefont {Yakushiji}, \citenamefont {Yuasa}, \citenamefont {Zvezdin}, \citenamefont {Ando},\ and\ \citenamefont {Fert}}]{dussaux_large_2010}%
  \BibitemOpen
  \bibfield  {author} {\bibinfo {author} {\bibfnamefont {A.}~\bibnamefont {Dussaux}}, \bibinfo {author} {\bibfnamefont {B.}~\bibnamefont {Georges}}, \bibinfo {author} {\bibfnamefont {J.}~\bibnamefont {Grollier}}, \bibinfo {author} {\bibfnamefont {V.}~\bibnamefont {Cros}}, \bibinfo {author} {\bibfnamefont {A.}~\bibnamefont {Khvalkovskiy}}, \bibinfo {author} {\bibfnamefont {A.}~\bibnamefont {Fukushima}}, \bibinfo {author} {\bibfnamefont {M.}~\bibnamefont {Konoto}}, \bibinfo {author} {\bibfnamefont {H.}~\bibnamefont {Kubota}}, \bibinfo {author} {\bibfnamefont {K.}~\bibnamefont {Yakushiji}}, \bibinfo {author} {\bibfnamefont {S.}~\bibnamefont {Yuasa}}, \bibinfo {author} {\bibfnamefont {K.}~\bibnamefont {Zvezdin}}, \bibinfo {author} {\bibfnamefont {K.}~\bibnamefont {Ando}},\ and\ \bibinfo {author} {\bibfnamefont {A.}~\bibnamefont {Fert}},\ }\bibfield  {title} {\bibinfo {title} {Large microwave generation from current-driven magnetic vortex oscillators in magnetic tunnel junctions},\ }\href
  {https://doi.org/10.1038/ncomms1006} {\bibfield  {journal} {\bibinfo  {journal} {Nature Communications}\ }\textbf {\bibinfo {volume} {1}},\ \bibinfo {pages} {8} (\bibinfo {year} {2010})}\BibitemShut {NoStop}%
\bibitem [{\citenamefont {Ivanov}\ and\ \citenamefont {Zaspel}(2007)}]{ivanov_excitation_2007}%
  \BibitemOpen
  \bibfield  {author} {\bibinfo {author} {\bibfnamefont {B.~A.}\ \bibnamefont {Ivanov}}\ and\ \bibinfo {author} {\bibfnamefont {C.~E.}\ \bibnamefont {Zaspel}},\ }\bibfield  {title} {\bibinfo {title} {Excitation of {Spin} {Dynamics} by {Spin}-{Polarized} {Current} in {Vortex} {State} {Magnetic} {Disks}},\ }\href {https://doi.org/10.1103/PhysRevLett.99.247208} {\bibfield  {journal} {\bibinfo  {journal} {Physical Review Letters}\ }\textbf {\bibinfo {volume} {99}},\ \bibinfo {pages} {247208} (\bibinfo {year} {2007})}\BibitemShut {NoStop}%
\bibitem [{\citenamefont {Kim}(2012)}]{kim_spintorque_2012}%
  \BibitemOpen
  \bibfield  {author} {\bibinfo {author} {\bibfnamefont {J.-V.}\ \bibnamefont {Kim}},\ }\bibfield  {title} {\bibinfo {title} {{Spin-torque oscillators}},\ }in\ \href {https://doi.org/10.1016/b978-0-12-397028-2.00004-7} {\emph {\bibinfo {booktitle} {Solid State Physics}}},\ Vol.~\bibinfo {volume} {63},\ \bibinfo {editor} {edited by\ \bibinfo {editor} {\bibfnamefont {R.~E.}\ \bibnamefont {Camley}}\ and\ \bibinfo {editor} {\bibfnamefont {R.~L.}\ \bibnamefont {Stamps}}}\ (\bibinfo  {publisher} {Academic Press},\ \bibinfo {address} {San Diego},\ \bibinfo {year} {2012})\ pp.\ \bibinfo {pages} {217 -- 294}\BibitemShut {NoStop}%
\bibitem [{\citenamefont {Khvalkovskiy}\ \emph {et~al.}(2010)\citenamefont {Khvalkovskiy}, \citenamefont {Grollier}, \citenamefont {Locatelli}, \citenamefont {Gorbunov}, \citenamefont {Zvezdin},\ and\ \citenamefont {Cros}}]{khvalkovskiy_nonuniformity_2010}%
  \BibitemOpen
  \bibfield  {author} {\bibinfo {author} {\bibfnamefont {A.~V.}\ \bibnamefont {Khvalkovskiy}}, \bibinfo {author} {\bibfnamefont {J.}~\bibnamefont {Grollier}}, \bibinfo {author} {\bibfnamefont {N.}~\bibnamefont {Locatelli}}, \bibinfo {author} {\bibfnamefont {Y.~V.}\ \bibnamefont {Gorbunov}}, \bibinfo {author} {\bibfnamefont {K.~A.}\ \bibnamefont {Zvezdin}},\ and\ \bibinfo {author} {\bibfnamefont {V.}~\bibnamefont {Cros}},\ }\bibfield  {title} {\bibinfo {title} {Nonuniformity of a planar polarizer for spin-transfer-induced vortex oscillations at zero field},\ }\href {https://doi.org/10.1063/1.3441405} {\bibfield  {journal} {\bibinfo  {journal} {Applied Physics Letters}\ }\textbf {\bibinfo {volume} {96}},\ \bibinfo {pages} {212507} (\bibinfo {year} {2010})}\BibitemShut {NoStop}%
\bibitem [{\citenamefont {Matsumoto}\ \emph {et~al.}(2019)\citenamefont {Matsumoto}, \citenamefont {Lequeux}, \citenamefont {Imamura},\ and\ \citenamefont {Grollier}}]{matsumoto_chaos_2019}%
  \BibitemOpen
  \bibfield  {author} {\bibinfo {author} {\bibfnamefont {R.}~\bibnamefont {Matsumoto}}, \bibinfo {author} {\bibfnamefont {S.}~\bibnamefont {Lequeux}}, \bibinfo {author} {\bibfnamefont {H.}~\bibnamefont {Imamura}},\ and\ \bibinfo {author} {\bibfnamefont {J.}~\bibnamefont {Grollier}},\ }\bibfield  {title} {\bibinfo {title} {Chaos and {Relaxation} {Oscillations} in {Spin}-{Torque} {Windmill} {Spiking} {Oscillators}},\ }\href {https://doi.org/10.1103/PhysRevApplied.11.044093} {\bibfield  {journal} {\bibinfo  {journal} {Physical Review Applied}\ }\textbf {\bibinfo {volume} {11}},\ \bibinfo {pages} {044093} (\bibinfo {year} {2019})}\BibitemShut {NoStop}%
\bibitem [{\citenamefont {Farcis}\ \emph {et~al.}(2023)\citenamefont {Farcis}, \citenamefont {Teixeira}, \citenamefont {Talatchian}, \citenamefont {Salomoni}, \citenamefont {Ebels}, \citenamefont {Auffret}, \citenamefont {Dieny}, \citenamefont {Mizrahi}, \citenamefont {Grollier}, \citenamefont {Sousa},\ and\ \citenamefont {Buda-Prejbeanu}}]{farcis_spiking_2023}%
  \BibitemOpen
  \bibfield  {author} {\bibinfo {author} {\bibfnamefont {L.}~\bibnamefont {Farcis}}, \bibinfo {author} {\bibfnamefont {B.~M.~S.}\ \bibnamefont {Teixeira}}, \bibinfo {author} {\bibfnamefont {P.}~\bibnamefont {Talatchian}}, \bibinfo {author} {\bibfnamefont {D.}~\bibnamefont {Salomoni}}, \bibinfo {author} {\bibfnamefont {U.}~\bibnamefont {Ebels}}, \bibinfo {author} {\bibfnamefont {S.}~\bibnamefont {Auffret}}, \bibinfo {author} {\bibfnamefont {B.}~\bibnamefont {Dieny}}, \bibinfo {author} {\bibfnamefont {F.~A.}\ \bibnamefont {Mizrahi}}, \bibinfo {author} {\bibfnamefont {J.}~\bibnamefont {Grollier}}, \bibinfo {author} {\bibfnamefont {R.~C.}\ \bibnamefont {Sousa}},\ and\ \bibinfo {author} {\bibfnamefont {L.~D.}\ \bibnamefont {Buda-Prejbeanu}},\ }\bibfield  {title} {\bibinfo {title} {Spiking {Dynamics} in {Dual} {Free} {Layer} {Perpendicular} {Magnetic} {Tunnel} {Junctions}},\ }\href {https://doi.org/10.1021/acs.nanolett.3c01597} {\bibfield  {journal} {\bibinfo  {journal} {Nano Letters}\ }\textbf {\bibinfo {volume}
  {23}},\ \bibinfo {pages} {7869} (\bibinfo {year} {2023})}\BibitemShut {NoStop}%
\bibitem [{\citenamefont {Taniguchi}(2019)}]{taniguchi_synchronized_2019}%
  \BibitemOpen
  \bibfield  {author} {\bibinfo {author} {\bibfnamefont {T.}~\bibnamefont {Taniguchi}},\ }\bibfield  {title} {\bibinfo {title} {{Synchronized, periodic, and chaotic dynamics in spin torque oscillator with two free layers}},\ }\href {https://doi.org/10.1016/j.jmmm.2019.03.090} {\bibfield  {journal} {\bibinfo  {journal} {Journal of Magnetism and Magnetic Materials}\ }\textbf {\bibinfo {volume} {483}},\ \bibinfo {pages} {281 } (\bibinfo {year} {2019})}\BibitemShut {NoStop}%
\bibitem [{\citenamefont {Cherepov}\ \emph {et~al.}(2012)\citenamefont {Cherepov}, \citenamefont {Koop}, \citenamefont {Galkin}, \citenamefont {Khymyn}, \citenamefont {Ivanov}, \citenamefont {Worledge},\ and\ \citenamefont {Korenivski}}]{cherepov_core-core_2012}%
  \BibitemOpen
  \bibfield  {author} {\bibinfo {author} {\bibfnamefont {S.~S.}\ \bibnamefont {Cherepov}}, \bibinfo {author} {\bibfnamefont {B.~C.}\ \bibnamefont {Koop}}, \bibinfo {author} {\bibfnamefont {A.~Y.}\ \bibnamefont {Galkin}}, \bibinfo {author} {\bibfnamefont {R.~S.}\ \bibnamefont {Khymyn}}, \bibinfo {author} {\bibfnamefont {B.~A.}\ \bibnamefont {Ivanov}}, \bibinfo {author} {\bibfnamefont {D.~C.}\ \bibnamefont {Worledge}},\ and\ \bibinfo {author} {\bibfnamefont {V.}~\bibnamefont {Korenivski}},\ }\bibfield  {title} {\bibinfo {title} {Core-{Core} {Dynamics} in {Spin} {Vortex} {Pairs}},\ }\href {https://doi.org/10.1103/PhysRevLett.109.097204} {\bibfield  {journal} {\bibinfo  {journal} {Physical Review Letters}\ }\textbf {\bibinfo {volume} {109}},\ \bibinfo {pages} {097204} (\bibinfo {year} {2012})}\BibitemShut {NoStop}%
\bibitem [{\citenamefont {Koop}\ \emph {et~al.}(2014)\citenamefont {Koop}, \citenamefont {Ivanov},\ and\ \citenamefont {Korenivski}}]{koop_2014_nonlin}%
  \BibitemOpen
  \bibfield  {author} {\bibinfo {author} {\bibfnamefont {B.~C.}\ \bibnamefont {Koop}}, \bibinfo {author} {\bibfnamefont {B.~A.}\ \bibnamefont {Ivanov}},\ and\ \bibinfo {author} {\bibfnamefont {V.}~\bibnamefont {Korenivski}},\ }\bibfield  {title} {\bibinfo {title} {Nonlinear dynamics in spin vortex pairs with strong core–core coupling},\ }\href {https://doi.org/10.1109/TMAG.2014.2327157} {\bibfield  {journal} {\bibinfo  {journal} {IEEE Transactions on Magnetics}\ }\textbf {\bibinfo {volume} {50}},\ \bibinfo {pages} {1} (\bibinfo {year} {2014})}\BibitemShut {NoStop}%
\bibitem [{\citenamefont {Bondarenko}\ \emph {et~al.}(2019)\citenamefont {Bondarenko}, \citenamefont {Holmgren}, \citenamefont {Li}, \citenamefont {Ivanov},\ and\ \citenamefont {Korenivski}}]{bondarenko_2019_chaos}%
  \BibitemOpen
  \bibfield  {author} {\bibinfo {author} {\bibfnamefont {A.~V.}\ \bibnamefont {Bondarenko}}, \bibinfo {author} {\bibfnamefont {E.}~\bibnamefont {Holmgren}}, \bibinfo {author} {\bibfnamefont {Z.~W.}\ \bibnamefont {Li}}, \bibinfo {author} {\bibfnamefont {B.~A.}\ \bibnamefont {Ivanov}},\ and\ \bibinfo {author} {\bibfnamefont {V.}~\bibnamefont {Korenivski}},\ }\bibfield  {title} {\bibinfo {title} {Chaotic dynamics in spin-vortex pairs},\ }\href {https://doi.org/10.1103/PhysRevB.99.054402} {\bibfield  {journal} {\bibinfo  {journal} {Phys. Rev. B}\ }\textbf {\bibinfo {volume} {99}},\ \bibinfo {pages} {054402} (\bibinfo {year} {2019})}\BibitemShut {NoStop}%
\bibitem [{\citenamefont {Locatelli}\ \emph {et~al.}(2011)\citenamefont {Locatelli}, \citenamefont {Naletov}, \citenamefont {Grollier}, \citenamefont {De~Loubens}, \citenamefont {Cros}, \citenamefont {Deranlot}, \citenamefont {Ulysse}, \citenamefont {Faini}, \citenamefont {Klein},\ and\ \citenamefont {Fert}}]{locatelli_dynamics_2011}%
  \BibitemOpen
  \bibfield  {author} {\bibinfo {author} {\bibfnamefont {N.}~\bibnamefont {Locatelli}}, \bibinfo {author} {\bibfnamefont {V.~V.}\ \bibnamefont {Naletov}}, \bibinfo {author} {\bibfnamefont {J.}~\bibnamefont {Grollier}}, \bibinfo {author} {\bibfnamefont {G.}~\bibnamefont {De~Loubens}}, \bibinfo {author} {\bibfnamefont {V.}~\bibnamefont {Cros}}, \bibinfo {author} {\bibfnamefont {C.}~\bibnamefont {Deranlot}}, \bibinfo {author} {\bibfnamefont {C.}~\bibnamefont {Ulysse}}, \bibinfo {author} {\bibfnamefont {G.}~\bibnamefont {Faini}}, \bibinfo {author} {\bibfnamefont {O.}~\bibnamefont {Klein}},\ and\ \bibinfo {author} {\bibfnamefont {A.}~\bibnamefont {Fert}},\ }\bibfield  {title} {\bibinfo {title} {Dynamics of two coupled vortices in a spin valve nanopillar excited by spin transfer torque},\ }\href {https://doi.org/10.1063/1.3553771} {\bibfield  {journal} {\bibinfo  {journal} {Applied Physics Letters}\ }\textbf {\bibinfo {volume} {98}},\ \bibinfo {pages} {062501} (\bibinfo {year} {2011})}\BibitemShut {NoStop}%
\bibitem [{\citenamefont {Sluka}\ \emph {et~al.}(2012)\citenamefont {Sluka}, \citenamefont {K{\'a}kay}, \citenamefont {Deac}, \citenamefont {B{\"u}rgler}, \citenamefont {Hertel},\ and\ \citenamefont {Schneider}}]{sluka_quenched_2012}%
  \BibitemOpen
  \bibfield  {author} {\bibinfo {author} {\bibfnamefont {V.}~\bibnamefont {Sluka}}, \bibinfo {author} {\bibfnamefont {A.}~\bibnamefont {K{\'a}kay}}, \bibinfo {author} {\bibfnamefont {A.~M.}\ \bibnamefont {Deac}}, \bibinfo {author} {\bibfnamefont {D.~E.}\ \bibnamefont {B{\"u}rgler}}, \bibinfo {author} {\bibfnamefont {R.}~\bibnamefont {Hertel}},\ and\ \bibinfo {author} {\bibfnamefont {C.~M.}\ \bibnamefont {Schneider}},\ }\bibfield  {title} {\bibinfo {title} {Quenched {Slonczewski} windmill in spin-torque vortex oscillators},\ }\href {https://doi.org/10.1103/PhysRevB.86.214422} {\bibfield  {journal} {\bibinfo  {journal} {Physical Review B}\ }\textbf {\bibinfo {volume} {86}},\ \bibinfo {pages} {214422} (\bibinfo {year} {2012})}\BibitemShut {NoStop}%
\bibitem [{\citenamefont {Hamadeh}\ \emph {et~al.}(2025)\citenamefont {Hamadeh}, \citenamefont {Koujok}, \citenamefont {Rodrigues}, \citenamefont {Riveros}, \citenamefont {Lomakin}, \citenamefont {Finocchio}, \citenamefont {De~Loubens}, \citenamefont {Klein},\ and\ \citenamefont {Pirro}}]{hamadeh_diverse_2025}%
  \BibitemOpen
  \bibfield  {author} {\bibinfo {author} {\bibfnamefont {A.~A.}\ \bibnamefont {Hamadeh}}, \bibinfo {author} {\bibfnamefont {A.}~\bibnamefont {Koujok}}, \bibinfo {author} {\bibfnamefont {D.~R.}\ \bibnamefont {Rodrigues}}, \bibinfo {author} {\bibfnamefont {A.}~\bibnamefont {Riveros}}, \bibinfo {author} {\bibfnamefont {V.}~\bibnamefont {Lomakin}}, \bibinfo {author} {\bibfnamefont {G.}~\bibnamefont {Finocchio}}, \bibinfo {author} {\bibfnamefont {G.}~\bibnamefont {De~Loubens}}, \bibinfo {author} {\bibfnamefont {O.}~\bibnamefont {Klein}},\ and\ \bibinfo {author} {\bibfnamefont {P.}~\bibnamefont {Pirro}},\ }\bibfield  {title} {\bibinfo {title} {Diverse dynamics in interacting vortices systems through tunable conservative and non-conservative coupling strengths},\ }\href {https://doi.org/10.1038/s42005-025-02006-3} {\bibfield  {journal} {\bibinfo  {journal} {Communications Physics}\ }\textbf {\bibinfo {volume} {8}},\ \bibinfo {pages} {85} (\bibinfo {year} {2025})}\BibitemShut {NoStop}%
\bibitem [{\citenamefont {Vansteenkiste}\ \emph {et~al.}(2014)\citenamefont {Vansteenkiste}, \citenamefont {Leliaert}, \citenamefont {Dvornik}, \citenamefont {Helsen}, \citenamefont {Garcia-Sanchez},\ and\ \citenamefont {Van~Waeyenberge}}]{vansteenkiste_design_2014}%
  \BibitemOpen
  \bibfield  {author} {\bibinfo {author} {\bibfnamefont {A.}~\bibnamefont {Vansteenkiste}}, \bibinfo {author} {\bibfnamefont {J.}~\bibnamefont {Leliaert}}, \bibinfo {author} {\bibfnamefont {M.}~\bibnamefont {Dvornik}}, \bibinfo {author} {\bibfnamefont {M.}~\bibnamefont {Helsen}}, \bibinfo {author} {\bibfnamefont {F.}~\bibnamefont {Garcia-Sanchez}},\ and\ \bibinfo {author} {\bibfnamefont {B.}~\bibnamefont {Van~Waeyenberge}},\ }\bibfield  {title} {\bibinfo {title} {The design and verification of {MuMax3}},\ }\href {https://doi.org/10.1063/1.4899186} {\bibfield  {journal} {\bibinfo  {journal} {AIP Advances}\ }\textbf {\bibinfo {volume} {4}},\ \bibinfo {pages} {107133} (\bibinfo {year} {2014})}\BibitemShut {NoStop}%
\bibitem [{\citenamefont {Hamadeh}\ \emph {et~al.}(2024)\citenamefont {Hamadeh}, \citenamefont {Koujok}, \citenamefont {Perna}, \citenamefont {Rodrigues}, \citenamefont {Riveros}, \citenamefont {Lomakin}, \citenamefont {Finocchio}, \citenamefont {De~Loubens}, \citenamefont {Klein},\ and\ \citenamefont {Pirro}}]{hamadeh_core_2024}%
  \BibitemOpen
  \bibfield  {author} {\bibinfo {author} {\bibfnamefont {A.}~\bibnamefont {Hamadeh}}, \bibinfo {author} {\bibfnamefont {A.}~\bibnamefont {Koujok}}, \bibinfo {author} {\bibfnamefont {S.}~\bibnamefont {Perna}}, \bibinfo {author} {\bibfnamefont {D.~R.}\ \bibnamefont {Rodrigues}}, \bibinfo {author} {\bibfnamefont {A.}~\bibnamefont {Riveros}}, \bibinfo {author} {\bibfnamefont {V.}~\bibnamefont {Lomakin}}, \bibinfo {author} {\bibfnamefont {G.}~\bibnamefont {Finocchio}}, \bibinfo {author} {\bibfnamefont {G.}~\bibnamefont {De~Loubens}}, \bibinfo {author} {\bibfnamefont {O.}~\bibnamefont {Klein}},\ and\ \bibinfo {author} {\bibfnamefont {P.}~\bibnamefont {Pirro}},\ }\bibfield  {title} {\bibinfo {title} {Core {Reversal} in {Vertically} {Coupled} {Vortices}: {Simulation} and {Experimental} {Study}},\ }\href {https://doi.org/10.1109/TNANO.2024.3420249} {\bibfield  {journal} {\bibinfo  {journal} {IEEE Transactions on Nanotechnology}\ }\textbf {\bibinfo {volume} {23}},\ \bibinfo {pages} {549} (\bibinfo {year}
  {2024})}\BibitemShut {NoStop}%
\bibitem [{\citenamefont {Van~Waeyenberge}\ \emph {et~al.}(2006)\citenamefont {Van~Waeyenberge}, \citenamefont {Puzic}, \citenamefont {Stoll}, \citenamefont {Chou}, \citenamefont {Tyliszczak}, \citenamefont {Hertel}, \citenamefont {F{\"a}hnle}, \citenamefont {Br{\"u}ckl}, \citenamefont {Rott}, \citenamefont {Reiss}, \citenamefont {Neudecker}, \citenamefont {Weiss}, \citenamefont {Back},\ and\ \citenamefont {Sch{\"u}tz}}]{van_waeyenberge_magnetic_2006}%
  \BibitemOpen
  \bibfield  {author} {\bibinfo {author} {\bibfnamefont {B.}~\bibnamefont {Van~Waeyenberge}}, \bibinfo {author} {\bibfnamefont {A.}~\bibnamefont {Puzic}}, \bibinfo {author} {\bibfnamefont {H.}~\bibnamefont {Stoll}}, \bibinfo {author} {\bibfnamefont {K.~W.}\ \bibnamefont {Chou}}, \bibinfo {author} {\bibfnamefont {T.}~\bibnamefont {Tyliszczak}}, \bibinfo {author} {\bibfnamefont {R.}~\bibnamefont {Hertel}}, \bibinfo {author} {\bibfnamefont {M.}~\bibnamefont {F{\"a}hnle}}, \bibinfo {author} {\bibfnamefont {H.}~\bibnamefont {Br{\"u}ckl}}, \bibinfo {author} {\bibfnamefont {K.}~\bibnamefont {Rott}}, \bibinfo {author} {\bibfnamefont {G.}~\bibnamefont {Reiss}}, \bibinfo {author} {\bibfnamefont {I.}~\bibnamefont {Neudecker}}, \bibinfo {author} {\bibfnamefont {D.}~\bibnamefont {Weiss}}, \bibinfo {author} {\bibfnamefont {C.~H.}\ \bibnamefont {Back}},\ and\ \bibinfo {author} {\bibfnamefont {G.}~\bibnamefont {Sch{\"u}tz}},\ }\bibfield  {title} {\bibinfo {title} {Magnetic vortex core reversal by excitation with short bursts
  of an alternating field},\ }\href {https://doi.org/10.1038/nature05240} {\bibfield  {journal} {\bibinfo  {journal} {Nature}\ }\textbf {\bibinfo {volume} {444}},\ \bibinfo {pages} {461} (\bibinfo {year} {2006})}\BibitemShut {NoStop}%
\bibitem [{\citenamefont {Bak}(1982)}]{bak_commensurate_1982}%
  \BibitemOpen
  \bibfield  {author} {\bibinfo {author} {\bibfnamefont {P.}~\bibnamefont {Bak}},\ }\bibfield  {title} {\bibinfo {title} {{Commensurate phases, incommensurate phases and the devil's staircase}},\ }\href {https://doi.org/10.1088/0034-4885/45/6/001} {\bibfield  {journal} {\bibinfo  {journal} {Reports on Progress in Physics}\ }\textbf {\bibinfo {volume} {45}},\ \bibinfo {pages} {587} (\bibinfo {year} {1982})}\BibitemShut {NoStop}%
\bibitem [{\citenamefont {Chaikin}\ and\ \citenamefont {Lubensky}(1995)}]{chaikin_principles_1995}%
  \BibitemOpen
  \bibfield  {author} {\bibinfo {author} {\bibfnamefont {P.~M.}\ \bibnamefont {Chaikin}}\ and\ \bibinfo {author} {\bibfnamefont {T.~C.}\ \bibnamefont {Lubensky}},\ }\href {https://doi.org/10.1017/CBO9780511813467} {\emph {\bibinfo {title} {{Principles of condensed matter physics}}}}\ (\bibinfo  {publisher} {Cambridge University Press},\ \bibinfo {address} {Cambridge, UK},\ \bibinfo {year} {1995})\BibitemShut {NoStop}%
\end{thebibliography}%

\end{document}